\documentclass[journal]{IEEEtran}
\usepackage{amsmath,amssymb,bm,graphicx,booktabs,array}
\usepackage[hidelinks,hypertexnames=false]{hyperref}
\usepackage{algorithm,algpseudocode}
\usepackage{capt-of}
\algrenewcommand\algorithmicrequire{\textbf{Input:}}
\algrenewcommand\algorithmicensure{\textbf{Output:}}
\graphicspath{{figures/}}
\newcommand{\HH}{^{\mathsf H}}
\newcommand{\TT}{^{\mathsf T}}
\newcommand{\E}{\mathbb E}
\newcommand{\CN}{\mathcal{CN}}
\newcommand{\diag}{\operatorname{diag}}
\newcommand{\Rea}{\operatorname{Re}}
\newcommand{\spx}{\operatorname{softplus}}
\newcommand{\sm}{\operatorname{softmax}}
\newcommand{\sig}{\operatorname{sigmoid}}
\newcommand{\norm}[1]{\left\lVert#1\right\rVert}

\newcommand{\NlosExpandedRows}{%
0 & -11.91 & -4.25 & -9.49 & -1.63 & 320.54 & 331.32 & 324.27 & 351.94 \\
5 & -16.25 & -10.63 & -13.10 & -8.40 & 211.07 & 224.10 & 216.11 & 233.43 \\
10 & -20.49 & -16.49 & -16.34 & -14.55 & 104.76 & 116.48 & 113.47 & 118.32 \\
15 & -25.44 & -22.35 & -18.96 & -20.69 & 27.18 & 33.54 & 39.99 & 31.93 \\
20 & -30.60 & -27.44 & -20.94 & -26.30 & 3.93 & 6.14 & 12.76 & 5.15 \\
25 & -33.88 & -30.88 & -22.44 & -29.73 & 0.88 & 1.46 & 6.36 & 1.64 \\
30 & -35.78 & -32.24 & -23.53 & -31.72 & 0.48 & 0.87 & 4.36 & 0.95 \\
}

\hypersetup{pdftitle={Learning-Assisted Channel Estimation for AFDM With Fractional Delay and Doppler},pdfauthor={Linyu Peng, Cunhua Pan, Hong Ren, Ruizhe Wang, Tian Qiu, Erkang Dong, Jiangzhou Wang}}
\begin{document}
\title{Learning-Assisted Channel Estimation for AFDM\\With Fractional Delay and Doppler}
\author{Linyu Peng, Cunhua Pan,~\IEEEmembership{Senior Member, IEEE}, Hong Ren,~\IEEEmembership{Member, IEEE},\protect\\%
Ruizhe Wang, Tian Qiu, Erkang Dong, and Jiangzhou Wang,~\IEEEmembership{Fellow, IEEE}%
\thanks{Linyu Peng, Cunhua Pan, Hong Ren, Ruizhe Wang, Tian Qiu, Erkang Dong, and Jiangzhou Wang are with National Mobile Communications Research Laboratory, Southeast University, Nanjing, China (e-mail: linyu.peng, cpan, hren, rzw, tianqiu, dek, j.z.wang@seu.edu.cn). Linyu Peng is also with Purple Mountain Laboratories, Nanjing, China.}}
\maketitle
\begin{abstract}
Affine frequency division multiplexing (AFDM) has emerged as a promising chirp-based waveform for high-mobility communications over doubly selective channels. However, channel estimation remains challenging in AFDM systems with fractional delay and Doppler, where each path response spreads across multiple discrete affine Fourier transform (DAFT) bins and overlaps with adjacent paths. Under a limited guard interval in an embedded-pilot frame, data leakage further degrades channel estimation. In this paper, we propose a learning-assisted continuous-parameter estimator (LaCE) that alternates between symbol inference and continuous multipath refinement from a pilot-based initialization. Specifically, each layer jointly fits the embedded pilot and reconstructed data responses via coupled damped Gauss--Newton delay--Doppler updates and regularized least-squares gain estimation. For low-complexity symbol inference, a gated recurrent unit approximates the symbol likelihoods directly from preconditioned conjugate-gradient iterates. Meanwhile, a joint neural network exploits complex AFDM correlations to calibrate symbol probabilities and adapt path-specific regularization and damping, mitigating data-feedback error propagation. Simulation results on 3GPP urban microcell (UMi) channels demonstrate that LaCE achieves lower normalized mean-square error (NMSE) and bit error rate (BER) than sparse-recovery, Bayesian, and continuous-parameter benchmarks across the considered signal-to-noise ratio (SNR) range, while maintaining reduced estimation latency.
\end{abstract}

\section{Introduction}
\IEEEPARstart{N}{ext-generation} wireless networks are expected to provide reliable broadband connectivity in high-mobility scenarios such as vehicular networks and high-speed railways \cite{bemani2023}. In these environments, multipath propagation and rapid relative motion induce significant delay and Doppler spreads, resulting in doubly selective channels. Within a transmission block, rapid time variation disrupts subcarrier orthogonality in conventional orthogonal frequency division multiplexing (OFDM)~\cite{zhu2009chunk,zhu2012chunk} and causes severe intercarrier interference. Against this background, affine frequency division multiplexing (AFDM) has emerged as a promising chirp-based multicarrier waveform for doubly selective channels \cite{bemani2023}. By modulating information symbols onto orthogonal chirps through the inverse discrete affine Fourier transform (DAFT), AFDM provides a one-dimensional representation of the doubly selective channel whose chirp parameters can be adapted to the channel delay and Doppler spreads \cite{bemani2023,zou2026afdm_parameters}. Consequently, AFDM can exploit full multipath diversity while facilitating applications to high-mobility multiple access and integrated sensing and communication \cite{luo2024scma,zhang2026afdm_pilot}.

The performance advantages of AFDM, however, rely critically on accurate channel state information (CSI) at the receiver. Under the idealized integer-delay model, an appropriately chosen chirp rate separates the DAFT-domain intervals associated with different delay taps, enabling full multipath diversity \cite{bemani2023}. In practical pulse-shaped systems, however, the physical path delays and Doppler shifts are generally not aligned with the discrete delay--Doppler grid. Specifically, a fractionally delayed path contributes to multiple discrete-time taps, each experiencing a distinct cyclic shift in the DAFT domain \cite{mirabella2026continuous,ni2026ambiguity}, while fractional Doppler further spreads each contribution across adjacent DAFT bins \cite{li2026mf}. As a result, responses from different propagation paths can strongly overlap, complicating the joint estimation of their continuous delays, Doppler shifts, and complex gains. Moreover, in an embedded-pilot frame, the same delay--Doppler spreading causes data responses to leak into the pilot observation region \cite{bemani2023,zhang2026afdm_pilot}. Although enlarging the guard interval can mitigate this leakage, it reduces spectral efficiency and does not resolve the mutual overlap among pilot responses. Sparse pilot designs have therefore been explored to curb guard overhead \cite{benzine2026sparse}.

Existing channel estimation methods for AFDM systems can be broadly categorized into grid-based and continuous-parameter approaches. Early embedded-pilot estimation in \cite{bemani2023} exploits the DAFT-domain separation of integer-delay taps, whereas diagonal reconstruction \cite{yin2024epadr} and sparse Bayesian learning (SBL) with adaptive Doppler dictionaries \cite{li2026iotsbl} further account for fractional Doppler. Nevertheless, these methods retain the integer-delay assumption and cannot capture the multi-tap dispersion induced by fractional delays. Extending sparse recovery to fractional delay and Doppler requires a two-dimensional dictionary over both parameters. While finer grid discretization reduces off-grid mismatch, it substantially increases computational complexity and dictionary coherence, making closely spaced paths difficult to distinguish.

Continuous-parameter estimation avoids grid discretization errors at the expense of nonlinear signal fitting. Matched filtering with generalized Fibonacci search (MF-GFS) lowers the search complexity by extracting one path at a time and refining its fractional Doppler over a local interval \cite{li2026mf}. However, its sequential extraction relies on approximate path orthogonality and is prone to error propagation when path responses overlap. Newtonized orthogonal matching pursuit (NOMP) and space-alternating generalized expectation-maximization (SAGE) alleviate error propagation by revisiting previously extracted paths during refinement \cite{nomp2016,wang2026sage}, yet repeated response evaluations incur substantial latency. In addition, the model-order-adaptive AFDM SAGE scheme in \cite{wang2026sage} leverages pilot phase differences across multiple blocks to enhance Doppler resolution, while other works employ multi-chirp pilot blocks \cite{zhou2026dualpilot} or multi-antenna and multi-block tensor decompositions \cite{wang2026tt} to decouple delay and Doppler. For pulse-shaped AFDM, ambiguity-function-assisted Newtonized channel estimation (AFNC) \cite{ma2026} refines fractional delays and Doppler shifts from a single pilot without data interference.

To further improve estimation accuracy without widening the pilot guard, data-aided schemes exploit observations from the data-bearing portion of the block while mitigating data-to-pilot interference. Under an integer-delay model, bilinear Gaussian belief propagation exchanges channel and symbol estimates in \cite{ranasinghe2025}. For superimposed-pilot AFDM systems, data-aided grid-evolution SBL (D-GESBL) feeds back detected symbols to iteratively update both the delay--Doppler dictionary and the posterior channel estimates \cite{luo2026gesbl}. Meanwhile, deep neural networks have been adopted to refine symbol moments in AFDM detection \cite{li2026afdm_ep} and to resolve closely spaced targets in AFDM sensing \cite{li2026afdm_sensing}. Nevertheless, extending data-aided processing to single-block continuous channel estimation remains challenging. Treating the predicted data interference as a fixed subtraction leaves data observations unused for refining path parameters, whereas jointly fitting pilots and reconstructed data becomes sensitive to symbol detection errors and high inter-path correlation.

Motivated by these challenges, we propose a learning-assisted continuous-parameter estimator (LaCE) for AFDM systems with embedded pilots under fractional delay and Doppler. Starting from a pilot-based initialization, LaCE unfolds alternating symbol inference and continuous path refinement into a small, fixed number of shared-weight layers, thereby bounding the per-block computational complexity and preventing the unpredictable iteration tails of conventional iterative solvers. Each layer jointly fits the pilot and reconstructed data responses via coupled damped Gauss--Newton delay--Doppler updates and regularized least-squares gain estimation, while a joint neural network leverages complex AFDM correlations to refine symbol probabilities and adapt path-specific gain penalties and damping, aided by a recurrent module that estimates symbol likelihoods directly from preconditioned conjugate-gradient (CG) iterates.

The main contributions of this paper are summarized as follows:
\begin{itemize}
\item We formulate a coupled continuous fit that jointly exploits the embedded pilots and reconstructed data. Within each layer, the reconstructed symbols remain fixed while their predicted response varies with the candidate delays, Doppler shifts, and gains. Retaining inter-path cross terms in the Gauss--Newton system explicitly accounts for overlapping responses.
\item A joint neural update and a CG-driven gated recurrent unit (CG--GRU) likelihood module are developed. CG--GRU estimates the symbol likelihood parameters directly from CG iterates to reduce inference complexity, while the joint network exploits complex AFDM correlations to calibrate symbol probabilities and adapt path-specific gain penalties and damping, suppressing error propagation from unreliable symbol feedback.
\item We analyze the per-block computational complexity and working memory, and validate the estimator on 3GPP urban microcell (UMi) line-of-sight (LOS) and non-line-of-sight (NLOS) channels. Simulation results show gains in normalized mean-square error (NMSE) and bit error rate (BER), with low estimation latency, while ablations isolate the gains of joint fitting, path-specific regularization, and CG--GRU approximation.
\end{itemize}

The remainder of this paper is organized as follows. Section~\ref{sec:model} presents the AFDM system model and problem formulation, Section~\ref{sec:estimator} develops LaCE along with its training and complexity analysis, Section~\ref{sec:simulations} provides simulation results, and Section~\ref{sec:conclusion} concludes the paper.

\textit{Notation:} Bold uppercase and lowercase symbols denote matrices and column vectors, respectively, while italic lowercase symbols denote scalars. For a matrix $\mathbf{X}$, $\mathbf{X}^*$, $\mathbf{X}\TT$, and $\mathbf{X}\HH$ denote its complex conjugate, transpose, and conjugate transpose, respectively. The Frobenius norm of $\mathbf{X}$ is denoted by $\norm{\mathbf{X}}_F$, and the $\ell_2$ norm of a vector $\mathbf{x}$ by $\norm{\mathbf{x}}$ or $\norm{\mathbf{x}}_2$. The $(m,n)$th entry of $\mathbf{X}$ is denoted by $[\mathbf{X}]_{m,n}$, and the $m$th entry of $\mathbf{x}$ by $[\mathbf{x}]_m$. The identity matrix is denoted by $\mathbf{I}$, or by $\mathbf{I}_n$ when its dimension $n$ is emphasized. The operator $\diag(\mathbf{x})$ forms a diagonal matrix with the entries of $\mathbf{x}$ on its main diagonal, while $\operatorname{tr}(\mathbf{X})$ and $\det(\mathbf{X})$ denote the trace and determinant, respectively. The operators $\E[\cdot]$ and $\operatorname{Cov}(\cdot)$ denote expectation and covariance, and $\Rea(\cdot)$ and $\operatorname{Im}(\cdot)$ return the real and imaginary parts, with $j=\sqrt{-1}$. The distribution $\CN(\mathbf{m},\mathbf{C})$ denotes a proper complex Gaussian random vector with mean $\mathbf{m}$ and covariance $\mathbf{C}$. A hat denotes an estimate, and a superscript $+$ denotes a quantity updated within a layer.

\section{System Model}
\label{sec:model}
Consider a single-input single-output (SISO) AFDM system operating over a doubly selective multipath channel, where each block comprises $N$ DAFT symbols partitioned into two known pilots, null guard symbols, and unknown data symbols. Excluding the chirp-periodic prefix and suffix, the block duration is $NT_s$, with $T_s$ denoting the sampling interval, and both the transmit pulse and its matched receive filter are known at the receiver. Within each block, the unknown number of physical paths $P$ and their continuous delays, Doppler frequencies, and complex gains remain constant and are estimated from the observations.

\subsection{AFDM Input--Output Model}
Let $\mathbf{x}=[x_0,\cdots,x_{N-1}]\TT\in\mathbb C^N$ be the DAFT-domain symbol vector, indexed by $\mathcal K=\{0,\cdots,N-1\}$. The unitary DAFT and its diagonal chirp matrices are
\begin{equation}
 \mathbf{A}=\bm{\Lambda}_{c_2}\mathbf{F}\bm{\Lambda}_{c_1},\qquad
 [\bm{\Lambda}_c]_{n,n}=e^{-j2\pi cn^2},
 \label{eq:daft}
\end{equation}
where $[\mathbf{F}]_{k,n}=N^{-1/2}e^{-j2\pi kn/N}$ and the real constants $c_1$ and $c_2$ are the AFDM chirp parameters \cite{bemani2023}. Applying the inverse DAFT (IDAFT) yields the time-domain sample vector $\mathbf{s}=\mathbf{A}\HH \mathbf{x}$:
\begin{equation}
 s[n]=\frac1{\sqrt N}\sum_{r=0}^{N-1}x_r
 e^{j2\pi(c_1n^2+c_2r^2+rn/N)},\quad n\in\mathcal K.
 \label{eq:idaft}
\end{equation}
Here $r$ indexes a transmitted DAFT symbol and $n$ indexes a time-domain sample. The chirp-periodic prefix and suffix are generated by
\begin{equation}
 s_e[m]=e^{j2\pi c_1(m^2-\bar m^2)}s[\bar m],\qquad
 \bar m=[m]_N,
 \label{eq:extension}
\end{equation}
where $[m]_N=m\bmod N$. The prefix and suffix provide the additional samples required by the channel delays and the finite pulse support. Joint design of the chirp rate and the time-domain guard duration can limit intersymbol and intercarrier interference \cite{zou2026afdm_parameters}. Let $q(t)$ be the real, unit-energy transmit pulse, with matched receive filter $q^*(-t)$. We form the transmit waveform by pulse-amplitude modulation of the extended IDAFT samples \cite{ni2026ambiguity}:
\begin{equation}
 s_{\mathrm{tx}}(t)=\sum_m s_e[m]q(t-mT_s).
 \label{eq:txwaveform}
\end{equation}
The sum includes the block samples and their prefix and suffix.

For path $i=1,\ldots,P$, let $g_i$, $\widetilde\tau_i$, and $f_i$ denote the complex gain, propagation delay, and Doppler frequency, respectively. We normalize the delay and Doppler as $\tau_i=\widetilde\tau_i/T_s$ and $\nu_i=-f_iNT_s$. The received waveform is
\begin{equation}
 r(t)=\sum_{i=1}^{P}g_i e^{-j2\pi\nu_it/(NT_s)}
 s_{\mathrm{tx}}(t-\tau_iT_s)+w_c(t).
 \label{eq:waveform}
\end{equation}
The Doppler phase in \eqref{eq:waveform} makes the channel time varying even though the path parameters are constant within each block. We consider fractional delay and Doppler with $0\leq\tau_i\leq\tau_{\max}$ and $|\nu_i|\leq\nu_{\max}$. The continuous-time noise $w_c(t)$ is circular white Gaussian noise. A constant path phase is included in $g_i$.

The receiver applies the matched filter $q^*(-t)$ to $r(t)$ and samples its output at $t=nT_s$ for $n=0,\ldots,N-1$. The samples are $\widetilde y_n=\int r(t)q^*(t-nT_s)\,dt$. Substituting \eqref{eq:txwaveform} and \eqref{eq:waveform} and setting $u=t-nT_s$ gives
\begin{align}
 \widetilde y_n={}&\sum_{i=1}^{P}g_i e^{-j2\pi\nu_i n/N}
 \sum_m s_e[m]\int q(u+(n-m-\tau_i)T_s)\nonumber\\
 &\hspace{28mm}\cdot q^*(u)e^{-j2\pi\nu_i u/(NT_s)}\,du+\widetilde w_n.
 \label{eq:matchedsample}
\end{align}
Let $\ell=n-m$ denote the discrete-time delay tap index. The effective pulse-response coefficient at tap $\ell$ is
\begin{equation}
 b_\ell(\tau,\nu)=\int q(u+(\ell-\tau)T_s)q^*(u)
 e^{-j2\pi\nu u/(NT_s)}\,du.
 \label{eq:pulse}
\end{equation}
For a pulse supported on $[-T_q,T_q]$, this coefficient is zero when $|\ell-\tau|T_s>2T_q$. Over $0\leq\tau\leq\tau_{\max}$, all contributing delay taps are contained in
\begin{equation}
 \mathcal I=\left\{\left\lceil-\frac{2T_q}{T_s}\right\rceil,\ldots,
 \left\lfloor\tau_{\max}+\frac{2T_q}{T_s}\right\rfloor\right\}.
 \label{eq:lagsupport}
\end{equation}
For $0\leq n\leq N-1$ and $\ell\in\mathcal I$, the transmit index $m=n-\ell$ lies between $-\ell_+$ and $N-1-\ell_-$, where $\ell_-=\min\mathcal I$ and $\ell_+=\max\mathcal I$. We choose the prefix and suffix to cover this range. With the coefficients in \eqref{eq:pulse}, \eqref{eq:matchedsample} becomes
\begin{equation}
 \widetilde y_n=\sum_{\ell\in\mathcal I}\bm{\Gamma}_{\ell,n}s_e[n-\ell]
 +\widetilde w_n,\quad n\in\mathcal K,
 \label{eq:samplechannel}
\end{equation}
where the time-varying channel impulse response is
\begin{equation}
 \bm{\Gamma}_{\ell,n}=\sum_{i=1}^{P}g_i b_\ell(\tau_i,\nu_i)e^{-j2\pi\nu_i n/N}.
 \label{eq:target}
\end{equation}
For a fractional delay, the pulse response can be nonzero across several discrete-time taps. Its coefficient $b_\ell(\tau,\nu)$ depends on both delay and Doppler \cite{ma2026}.

For a unit-gain path at $\bm{\theta}=[\tau,\nu]\TT$, the matrix $\bm{\mathcal T}(\bm{\theta})\in\mathbb C^{N\times N}$ maps the transmitted DAFT symbol vector $\mathbf{x}$ to that path's contribution to the $N$ matched-filter samples of the block:
\begin{equation}
 [\bm{\mathcal T}(\bm{\theta})\mathbf{x}]_n=e^{-j2\pi\nu n/N}
 \sum_{\ell\in\mathcal I}b_\ell(\tau,\nu)s_e[n-\ell].
 \label{eq:operator}
\end{equation}
Substituting the IDAFT samples from \eqref{eq:idaft} into \eqref{eq:extension} gives
\begin{equation}
 s_e[n-\ell]=\frac1{\sqrt N}\sum_{r=0}^{N-1}x_r
 e^{j2\pi\{c_1(n-\ell)^2+c_2r^2+r(n-\ell)/N\}}.
 \label{eq:extendedidaft}
\end{equation}
Applying the DAFT to \eqref{eq:operator} cancels the quadratic terms in $n$. The remaining phase in $n$ is $-2\pi(k-r+2Nc_1\ell+\nu)n/N$. Consequently, $\mathbf{K}(\bm{\theta})=\mathbf{A}\bm{\mathcal T}(\bm{\theta})$ has entries
\begin{align}
 [\mathbf{K}(\bm{\theta})]_{k,r}={}&e^{j2\pi c_2(r^2-k^2)}
 \sum_{\ell\in\mathcal I}b_\ell(\tau,\nu)
 e^{j2\pi(c_1\ell^2-r\ell/N)}\nonumber\\
 &\hspace{14mm}\cdot D_N(k-r+2Nc_1\ell+\nu),
 \label{eq:afdmresponse}\\
 D_N(\omega)={}&\frac1N\sum_{n=0}^{N-1}e^{-j2\pi\omega n/N}.
 \label{eq:dirichlet}
\end{align}
Here $k$ and $r$ index received and transmitted DAFT bins, respectively. Applying the DAFT to the $N$ matched-filter samples $\widetilde{\mathbf{y}}$ gives
\begin{equation}
 \mathbf{y}_{\mathrm A}=\mathbf{H}_{\mathrm A}\mathbf{x}+\mathbf{w}_{\mathrm A},\qquad
 \mathbf{H}_{\mathrm A}=\sum_{i=1}^{P}g_i\mathbf{K}(\bm{\theta}_i),
 \label{eq:afdmchannel}
\end{equation}
where $\mathbf{y}_{\mathrm A}=\mathbf{A}\widetilde{\mathbf{y}}$ and $\mathbf{w}_{\mathrm A}=\mathbf{A}\widetilde{\mathbf{w}}$.

In \eqref{eq:afdmresponse}, the response associated with tap $\ell$ is centered at $k=r-2Nc_1\ell-\nu$. In the single-tap, integer-delay limit, choosing integer $2Nc_1>2\nu_{\max}$ separates the Doppler intervals associated with adjacent delay bins \cite{bemani2023}. With fractional delay, contributions from multiple taps can overlap in the same DAFT bin. The chirp parameter $c_2$ contributes the phase factor shown in \eqref{eq:afdmresponse}.

The noise component of $\widetilde y_n$ is
\[
 \widetilde w_n=\int w_c(t)q^*(t-nT_s)\,dt.
\]
The covariance of the matched-filter noise is $\operatorname{Cov}(\widetilde{\mathbf{w}})=\sigma^2\mathbf{C}_q$, where $[\mathbf{C}_q]_{nm}=\int q^*(t-nT_s)q(t-mT_s)\,dt$. Filtering can correlate the noise samples even when $w_c(t)$ is white. In particular, two samples are correlated when the corresponding shifted pulses have a nonzero inner product.

To whiten the sampled noise, we compute the Cholesky factorization $\mathbf{C}_q=\mathbf{L}_q\mathbf{L}_q\HH$. Multiplication by $\bm{\mathcal R}=\mathbf{L}_q^{-1}$ gives $\mathbf{y}=\bm{\mathcal R}\widetilde{\mathbf{y}}$ and $\mathbf{T}(\bm{\theta})=\bm{\mathcal R}\bm{\mathcal T}(\bm{\theta})$. The resulting observation model is
\begin{equation}
 \mathbf{y}=\mathbf{H}(\mathbf{Q},\mathbf{g})\mathbf{x}+\mathbf{w},\qquad \mathbf{H}(\mathbf{Q},\mathbf{g})=\sum_{i=1}^{P}g_i\mathbf{T}(\bm{\theta}_i),
 \label{eq:fullchannel}
\end{equation}
where $\mathbf{Q}=[\bm{\theta}_1,\ldots,\bm{\theta}_P]$, $\mathbf{g}=[g_1,\ldots,g_P]\TT$, and $\mathbf{w}=\bm{\mathcal R}\widetilde{\mathbf{w}}\sim\CN(\mathbf{0},\sigma^2\mathbf{I})$. Since $\mathbf{y}_{\mathrm A}=\mathbf{A}\widetilde{\mathbf{y}}$ and $\mathbf{y}=\bm{\mathcal R}\widetilde{\mathbf{y}}$, their channel matrices satisfy $\mathbf{H}(\mathbf{Q},\mathbf{g})=\bm{\mathcal R} \mathbf{A}\HH \mathbf{H}_{\mathrm A}$. We use the DAFT-domain model in \eqref{eq:afdmchannel} to describe data leakage into the pilot observations and the whitened time-domain model in \eqref{eq:fullchannel} for estimation. Channel estimation accuracy is evaluated in terms of the impulse response $\bm{\Gamma}$ in \eqref{eq:target}.

\subsection{Embedded Pilots and Observation}
Partition the transmit DAFT bins into pilot, guard, and data bins, with index sets $\mathcal P$, $\mathcal Z$, and $\mathcal D$, respectively. The two pilot bins $k_1$ and $k_2$ lie in the contiguous interval $\mathcal G$. The other bins in $\mathcal G$ form $\mathcal Z=\mathcal G\setminus\mathcal P$, and the data bins form $\mathcal D=\mathcal K\setminus\mathcal G$. The transmitted block is
\begin{align}
 \mathbf{x}&=\mathbf{p}+\mathbf{S}\mathbf{d},\qquad \mathbf{S}=[\mathbf{e}_{\mathcal D_1},\ldots,\mathbf{e}_{\mathcal D_D}],
 \label{eq:source}\\
 \mathbf{p}&=\sqrt{E_p/2}(\mathbf{e}_{k_1}+e^{j\varphi}\mathbf{e}_{k_2}),
 \label{eq:pilotlayout}
\end{align}
where $\mathbf{e}_k$ is the $k$th canonical vector, $E_p$ is the total pilot energy, and $\varphi$ is the known relative phase. The data vector $\mathbf{d}\in\mathcal A^D$ contains $D=|\mathcal D|$ independent zero-mean symbols from the modulation constellation $\mathcal A$, with average energy $E_s$.

From \eqref{eq:afdmresponse}, the response of transmit bin $r$ is centered near $r-2Nc_1\ell-\nu$ for each delay tap $\ell$. Let $\Omega$ be a set of receive DAFT bins around the pilot responses, and let $\mathbf{R}_\Omega$ select those bins. The expected data leakage energy into $\Omega$ through a unit-gain path is
\begin{equation}
 \E_{\mathbf{d}}\norm{\mathbf{R}_\Omega \mathbf{K}(\bm{\theta})\mathbf{S}\mathbf{d}}^2
 =E_s\sum_{r\in\mathcal D}\norm{\mathbf{R}_\Omega \mathbf{K}(\bm{\theta})\mathbf{e}_r}^2.
 \label{eq:guardenergy}
\end{equation}
The guard limits data leakage into $\Omega$ by setting selected transmit symbols to zero. Under fractional Doppler, the responses of data symbols outside the guard may still extend into $\Omega$. The relative phase of the two pilot responses varies with $\ell$ through the factor $e^{-j2\pi(k_2-k_1)\ell/N}$ in \eqref{eq:afdmresponse}. For initialization, the receiver evaluates their combined response $\mathbf{T}(\bm{\theta})\mathbf{p}$ over the entire received block. Section~\ref{sec:simulations} specifies the pilot locations and guard allocation.

Substitution of \eqref{eq:source} into \eqref{eq:fullchannel} gives
\begin{equation}
 \mathbf{y}=\sum_{i=1}^{P}g_i\mathbf{T}(\tau_i,\nu_i)(\mathbf{p}+\mathbf{S}\mathbf{d})+\mathbf{w}.
 \label{eq:model}
\end{equation}
Given the whitened observation $\mathbf{y}$ in \eqref{eq:model} along with the known pulse shape, chirp parameters, frame allocation, constellation $\mathcal A$, and delay--Doppler search region, the objective is to recover the time-varying impulse response $\bm{\Gamma}$ while jointly inferring the unknown data symbols $\mathbf{d}$, without prior knowledge of the physical path count $P$, noise variance $\sigma^2$, or nominal signal-to-noise ratio (SNR).

\section{Proposed LaCE Framework}
\label{sec:estimator}

\begin{figure*}[!t]
\centering\includegraphics[width=\textwidth]{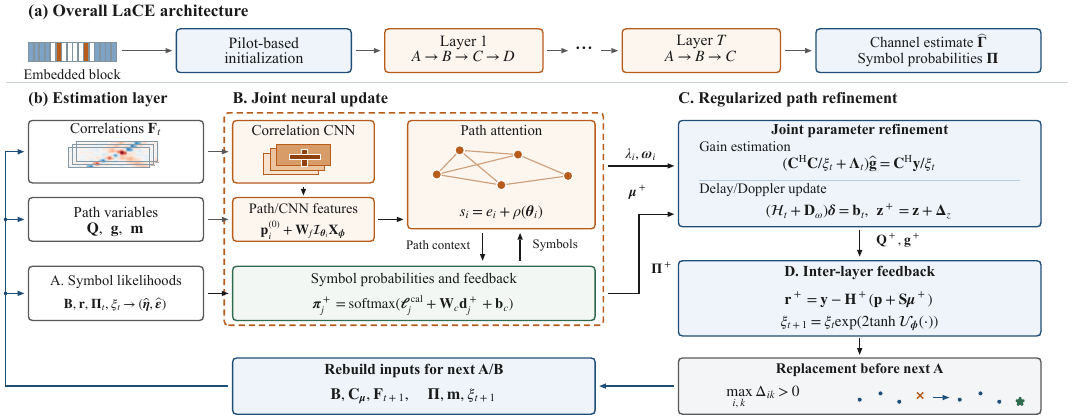}
\caption{Overview of LaCE: (a) shared-weight estimation layers; (b) operations described in Sections~\ref{sec:cavitylearning}--\ref{sec:replacementschedule}.}
\label{fig:receiver}
\end{figure*}

Figure~\ref{fig:receiver} gives an overview of LaCE. LaCE represents the channel by $V$ \emph{candidate paths}, each parameterized by a continuous delay--Doppler coordinate vector $\bm{\theta}_i=[\tau_i,\nu_i]\TT$ and complex gain $g_i$. Here $V$ is a preset over-parameterized candidate count, and inactive candidates are suppressed by learned path-specific gain regularization. Starting from a pilot-based initialization $(\mathbf{Q}_1,\mathbf{g}_1)$, LaCE unfolds estimation into $T$ shared-weight layers, where $\bm{\phi}$ collects all learnable neural network parameters across the framework.

Within each layer $t$, after an optional path-replacement test for $t>1$, LaCE executes four stages: (A) low-complexity symbol likelihood evaluation via $K_c$ CG steps and a GRU (Section~\ref{sec:cavitylearning}); (B) a joint neural update that refines symbol probabilities and predicts path-specific gain regularization and coordinate damping (Section~\ref{sec:interpretation}); (C) continuous path refinement via $J$ coupled damped Gauss--Newton steps and regularized least-squares gain fitting (Section~\ref{sec:pathrecovery}); and (D) inter-layer noise-variance feedback and state handoff for $t<T$ (Section~\ref{sec:replacementschedule}). A discrete grid $\{\bm{\vartheta}_k\}_{k=1}^{G}$ is used solely for initialization and path replacement, whereas all refinement operates over continuous coordinates.

At layer $t$, the path-coordinate matrix $\mathbf{Q}_t=[\bm{\theta}_{1,t},\ldots,\bm{\theta}_{V,t}]$ and gain vector $\mathbf{g}_t$ define the channel matrix $\mathbf{H}_t=\sum_{i=1}^V g_{i,t}\mathbf{T}(\bm{\theta}_{i,t})$ and impulse response estimate $\widehat{\bm{\Gamma}}_t$ via \eqref{eq:target}. For each unknown data symbol $d_j$, LaCE maintains constellation probability vectors $\bm{\pi}_j$ with entries $\pi_j(a)$ ($a\in\mathcal A$), stacked into $\bm{\Pi}=[\bm{\pi}_1,\ldots,\bm{\pi}_D]$ with mean and variance
\begin{equation}
 \mu_j=\sum_{a\in\mathcal A}a\pi_j(a),\qquad
 v_j=\sum_{a\in\mathcal A}|a|^2\pi_j(a)-|\mu_j|^2.
 \label{eq:moments}
\end{equation}
Collecting $\bm{\mu}=[\mu_1,\ldots,\mu_D]\TT$ and $\mathbf{v}=[v_1,\ldots,v_D]\TT$, the reconstructed DAFT symbol vector is $\bar{\mathbf{x}}=\mathbf{p}+\mathbf{S}\bm{\mu}$. We define the path response atoms $\mathbf{C}_{\bm{\mu}}(\mathbf{Q})$, the effective data channel $\mathbf{B}$, and the received data-interference covariance $\mathbf{R}_{\mathrm{data}}$ as
\begin{align}
 \mathbf{a}_{\bm{\mu}}(\bm{\theta})&=\mathbf{T}(\bm{\theta})\bar{\mathbf{x}},\nonumber\\
 \mathbf{C}_{\bm{\mu}}(\mathbf{Q})&=[\mathbf{a}_{\bm{\mu}}(\bm{\theta}_1),\ldots,\mathbf{a}_{\bm{\mu}}(\bm{\theta}_V)],
 \label{eq:atoms}\\
 \mathbf{B}&=\mathbf{H}(\mathbf{Q},\mathbf{g})\mathbf{S},\qquad \mathbf{R}_{\mathrm{data}}=\mathbf{B}\diag(\mathbf{v})\mathbf{B}\HH.
 \label{eq:datacov}
\end{align}
Accordingly, path refinement fits the predicted received signal $\mathbf{H}\bar{\mathbf{x}}=\mathbf{C}_{\bm{\mu}}(\mathbf{Q})\mathbf{g}$, while symbol inference operates on $\mathbf{y}=\mathbf{H}\mathbf{p}+\mathbf{B}\mathbf{d}+\mathbf{w}$, where the diagonal entries $[\mathbf{v}_y]_n=[\mathbf{R}_{\mathrm{data}}]_{n,n}=\sum_j|[\mathbf{B}]_{n,j}|^2v_j$ give the per-sample data-interference variance (known pilots and guards contribute zero variance).

To initialize layer 1 from $\mathbf{y}$, LaCE sets uniform symbol probabilities $\pi_j(a)=1/|\mathcal A|$, unit auxiliary gain scales $m_i=1$, and zero hidden states $\mathbf{H}_p,\mathbf{H}_d$, and predicts the initial effective noise variance
\begin{equation}
 \xi_1=P_y\exp\!\left(\mathcal A_{\bm{\phi}}(\mathbf{f}_0(\mathbf{y}))\right),\qquad P_y=\norm{\mathbf{y}}^2/N,
 \label{eq:initialscale}
\end{equation}
where $\mathbf{f}_0(\mathbf{y})$ collects normalized received-power statistics and sample correlations. Initial candidate locations are obtained by a sequential pursuit search with joint gain re-estimation \cite{nomp2016}: starting from $\mathbf{r}_0=\mathbf{y}$, each unused grid point $\bm{\vartheta}_k$ is scored by $|\mathbf{a}_p(\bm{\vartheta}_k)\HH \mathbf{r}|^2/\norm{\mathbf{a}_p(\bm{\vartheta}_k)}^2$ with pilot response $\mathbf{a}_p(\bm{\vartheta}_k)=\mathbf{T}(\bm{\vartheta}_k)\mathbf{p}$. At each step, the highest-scoring point is added, all selected gains are re-estimated via regularized least squares, and the residual is updated until $V$ candidates are chosen. The selected coordinates are then refined through two pilot-only Gauss--Newton steps (each followed by joint gain re-estimation), using fixed sample weights $(\xi_1+[\mathbf{v}_y]_n)^{-1}$ evaluated under uniform $\bm{\Pi}$ and fixed gain penalties computed from the initial gains and weighted response energies. Any grid point on the search boundary is shifted slightly into the cell interior prior to unconstrained coordinate mapping, producing $(\mathbf{Q}_1,\mathbf{g}_1)$ for layer 1.

\subsection{Symbol Likelihoods}
\label{sec:cavitylearning}
For the current channel $\mathbf{H}=\mathbf{H}(\mathbf{Q},\mathbf{g})$, subtracting the pilot contribution $\mathbf{b}_p=\mathbf{H}\mathbf{p}$ gives $\mathbf{y}-\mathbf{b}_p=\mathbf{B}\mathbf{d}+\mathbf{w}$, where $\mathbf{B}=\mathbf{H}\mathbf{S}=[\mathbf{b}_1,\ldots,\mathbf{b}_D]$ is the data channel. Subtracting the predicted data contribution $\mathbf{B}\bm{\mu}$ gives the residual $\mathbf{r}=\mathbf{y}-\mathbf{b}_p-\mathbf{B}\bm{\mu}=\mathbf{y}-\mathbf{C}_{\bm{\mu}}(\mathbf{Q})\mathbf{g}$. Using Gaussian soft interference cancellation \cite{ranasinghe2025}, we subtract the mean contributions of the other symbols and model their residual interference as Gaussian. This gives a likelihood for $d_j$ that depends on the other symbols' priors but excludes the prior on $d_j$:
\begin{align}
 \mathbf{R}_{-j}&=\xi_t\mathbf{I}+\sum_{k\ne j}v_k\mathbf{b}_k\mathbf{b}_k\HH,\nonumber\\
 \mathbf{r}_{-j}&=\mathbf{y}-\mathbf{b}_p-\sum_{k\ne j}\mathbf{b}_k\mu_k,\nonumber\\
 \epsilon_j&=\mathbf{b}_j\HH \mathbf{R}_{-j}^{-1}\mathbf{b}_j,\qquad
 \eta_j=\mathbf{b}_j\HH \mathbf{R}_{-j}^{-1}\mathbf{r}_{-j}.
 \label{eq:cavity}
\end{align}
Up to an additive constant, the log-likelihood of $d_j=a$ is $2\Rea(a^*\eta_j)-\epsilon_j|a|^2$, where $\epsilon_j$ and $\eta_j$ denote the precision and precision-weighted mean, respectively. To share the computation across symbols, set $\mathbf{S}_v=\diag(\sqrt{\mathbf{v}})$ and solve a single linear system:
\begin{align}
 \bm{\mathsf K}\bm{\chi}&=\mathbf{S}_v\mathbf{B}\HH \mathbf{r}/\xi_t,\nonumber\\
 \bm{\mathsf K}&=\mathbf{I}+\mathbf{S}_v\mathbf{B}\HH \mathbf{B}\mathbf{S}_v/\xi_t.
 \label{eq:pcg}
\end{align}
Let $\mathbf{q}_{\mathrm{CG}}=\mathbf{S}_v\mathbf{B}\HH \mathbf{r}/\xi_t$. For the exact solution $\bm{\chi}$ of \eqref{eq:pcg}, the Gaussian posterior mean is $\bm{\mu}+\mathbf{S}_v\bm{\chi}$. The matrix $\bm{\mathsf K}$ is Hermitian positive definite. We use its diagonal as the preconditioner $\mathbf{M}_c$ and apply $\bm{\mathsf K}$ to any vector $\bm{\psi}\in\mathbb C^D$ via matrix--vector products with $\mathbf{B}$ and $\mathbf{B}\HH$:
\begin{align}
 [\mathbf{M}_c]_{jj}&=1+v_j\norm{\mathbf{b}_j}^2/\xi_t,\nonumber\\
 \bm{\mathsf K}\bm{\psi}&=\bm{\psi}+\mathbf{S}_v\mathbf{B}\HH(\mathbf{B}\mathbf{S}_v\bm{\psi})/\xi_t.
 \label{eq:pcgoperator}
\end{align}
Each CG step therefore requires one multiplication by $\mathbf{B}$ and one by $\mathbf{B}\HH$. The diagonal entries of the preconditioner depend on each symbol variance and the energy of its channel column.

Starting from $\bm{\chi}^{(0)}=\mathbf{0}$, we run $K_c$ preconditioned CG steps \cite{fang2026cg} and denote the residual at step $s$ by $\mathbf{r}_c^{(s)}=\mathbf{q}_{\mathrm{CG}}-\bm{\mathsf K}\bm{\chi}^{(s)}$.

To express the likelihood in terms of the linear system, let $\kappa_j=[\bm{\mathsf K}^{-1}]_{jj}$ and $\mathbf{u}=\mathbf{S}_v^{-1}\bm{\chi}$ for the exact solution. Applying the Woodbury matrix identity and rank-one update yields
\begin{equation}
 \epsilon_j=\frac{1-\kappa_j}{v_j\kappa_j},\qquad
 z_j\equiv\frac{\eta_j}{\epsilon_j}
 =\mu_j+u_j(v_j+\epsilon_j^{-1}).
 \label{eq:extrinsic}
\end{equation}
Although solving \eqref{eq:pcg} via CG avoids forming $\bm{\mathsf K}^{-1}$ when computing $\mathbf{u}$, evaluating the exact likelihood precision $\epsilon_j$ still requires the diagonal entries $\kappa_j=[\bm{\mathsf K}^{-1}]_{jj}$, which incurs costly $O(D^3)$ matrix inversion.

\begin{figure*}[!t]
\centering\includegraphics[width=0.95\textwidth]{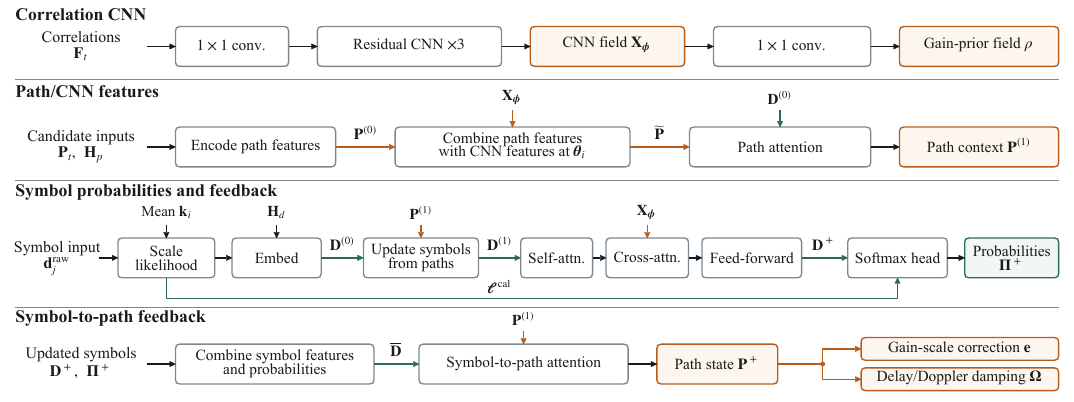}
\caption{Joint neural update (block B in Fig.~\ref{fig:receiver}). Complex AFDM correlations, candidate features, and symbol feedback contribute to the path-state update.}
\label{fig:jointnetwork}
\end{figure*}

To avoid this matrix inversion, a GRU with 48 hidden units \cite{cho2014} approximates the likelihood parameters directly from the CG iterates. Its weights are shared across symbols and steps. For symbol $j$ at step $s$, its 29-dimensional real input vector $\bm{\varphi}_j^{(s)}$ includes $\bm{\pi}_j$, $\mu_j$, $v_j$, the mean correction $\delta\mu_j^{(s)}$, and the scaled residual $\widetilde r_{c,j}^{(s)}$. The remaining inputs describe the precision bound, preconditioner diagonal, effective noise variance, step and layer indices, and residual-to-noise power:
\begin{align}
 \delta\mu_j^{(s)}&=[\mathbf{S}_v\bm{\chi}^{(s)}]_j,\qquad
 \widetilde r_{c,j}^{(s)}=[\mathbf{M}_c^{-1/2}\mathbf{r}_c^{(s)}]_j,\nonumber\\
 \mathbf{h}_{c,j}^{(s)}&=\operatorname{GRU}_{\bm{\phi}}(\bm{\varphi}_j^{(s)},\mathbf{h}_{c,j}^{(s-1)}),\nonumber\\
 [\Rea c_j,\operatorname{Im}c_j,a_j]\TT&=\mathbf{W}_o\mathbf{h}_{c,j}^{(K_c)}+\mathbf{b}_o.
 \label{eq:grureadout}
\end{align}
Each likelihood calculation starts from $\mathbf{h}_{c,j}^{(0)}=\mathbf{0}$. A linear projection of the final GRU state gives a real parameter $a_j$ for the precision and a complex correction $c_j$ for the mean. With $\widehat{\mathbf{u}}=\mathbf{S}_v^{-1}\bm{\chi}^{(K_c)}$, the likelihood parameters are
\begin{align}
 \widehat\epsilon_j&=\frac{\norm{\mathbf{b}_j}^2}{\xi_t}\sig(a_j),\nonumber\\
 \widehat z_j&=\mu_j+\widehat u_j(v_j+\widehat\epsilon_j^{-1})
 +c_j\sqrt{v_j+\widehat\epsilon_j^{-1}}.
 \label{eq:learnedcavity}
\end{align}
The sigmoid keeps the precision between zero and $\norm{\mathbf{b}_j}^2/\xi_t$. Together, $\widehat\epsilon_j$ and $\widehat\eta_j=\widehat\epsilon_j\widehat z_j$ determine the log-likelihood passed to the joint network.

\subsection{Joint Neural Update}
\label{sec:interpretation}

The symbol likelihoods in Section~\ref{sec:cavitylearning} rely on the current channel estimate $\mathbf{H}=\mathbf{H}(\mathbf{Q},\mathbf{g})$, so channel estimation errors directly affect soft symbol inference. Letting $\mathbf{H}_\star$ denote the true channel, the residual decomposes as
\begin{equation}
 \mathbf{r}=\underbrace{(\mathbf{H}_\star-\mathbf{H})(\mathbf{p}+\mathbf{S}\mathbf{d})}_{\text{channel mismatch}}
 +\mathbf{B}(\mathbf{d}-\bm{\mu})+\mathbf{w},
 \label{eq:fielddecomp}
\end{equation}
which couples channel mismatch, symbol errors, and noise. Consequently, the residual norm alone cannot indicate which candidate paths suffer from strong overlap or require tighter regularization. To expose this structure over the delay--Doppler plane, we evaluate the response $\mathbf{f}_k=\mathbf{a}_{\bm{\mu}}(\bm{\vartheta}_k)$ at each grid point $\bm{\vartheta}_k$ ($k=1,\ldots,G$) and compute the normalized residual and received-signal correlations
\begin{equation}
 u_k=\frac{\mathbf{f}_k\HH \mathbf{r}}{\sqrt{\xi_t\norm{\mathbf{f}_k}^2}},\qquad
 t_k=\frac{\mathbf{f}_k\HH \mathbf{y}}{\sqrt{\xi_t\norm{\mathbf{f}_k}^2}}.
 \label{eq:field}
\end{equation}
The real and imaginary parts of $(u_k,t_k)$, together with response energies and grid coordinates, form the 2-D correlation tensor $\mathbf{F}_t$. Note that the response inner product $\mathcal K_{\bm{\mu}}(\bm{\theta},\bm{\theta}')=\mathbf{a}_{\bm{\mu}}(\bm{\theta})\HH \mathbf{a}_{\bm{\mu}}(\bm{\theta}')$ quantifies inter-path overlap and forms the Gram matrix when evaluated across candidate pairs. Because $\mathbf{a}_{\bm{\mu}}$ depends on $\bm{\mu}$, refining the data estimates simultaneously sharpens the correlation features and the subsequent path fit.

As illustrated in Fig.~\ref{fig:jointnetwork}, the joint network $\mathcal N_{\bm{\phi}}$ integrates these AFDM correlations with path and symbol states via
\begin{equation}
 (\mathbf{e},\rho,\bm{\Omega},\bm{\Pi}^+,\mathbf{H}_p^+,\mathbf{H}_d^+)=\mathcal N_{\bm{\phi}}(\mathcal E_t,\mathbf{H}_p,\mathbf{H}_d),
 \label{eq:network}
\end{equation}
where $\mathbf{H}_p,\mathbf{H}_d$ stack the previous-layer hidden vectors $\mathbf{h}_{p,i},\mathbf{h}_{d,j}$, $\mathbf{e}=[e_1,\ldots,e_V]\TT$ and $\bm{\Omega}=[\bm{\omega}_1,\ldots,\bm{\omega}_V]\TT$ collect the gain-scale adjustments and damping vectors, and $\mathcal E_t=(\mathbf{F}_t,\bm{\mathsf P}_t,\bm{\mathsf D}_t,\bm{\mathsf R})$. Here, $\bm{\mathsf P}_t$ encodes each candidate's coordinates, gain statistics, scale $m_i$, and local curvature $\mathbf{k}_i$, while $\bm{\mathsf D}_t$ encodes symbol probabilities, moments, and calibrated log-likelihoods, and $\bm{\mathsf R}$ encodes relative path coordinates, symbol entropy, and symbol variance. Specifically, for each candidate path we form a local regularized Gauss--Newton matrix under $\bm{\mu}$ and extract the four-element vector $\mathbf{k}_i$ from the diagonal entries of its inverse as approximate local error variances, rescaling the two gain entries to $g_i/m_i$ before log-transformation. Matching the regular 2-D correlation grid and irregular path--symbol structure, $\mathcal N_{\bm{\phi}}$ combines a lightweight separable convolutional encoder, path--symbol attention, and a shared calibration head with one compact parameter set $\bm{\phi}$ across all layers.

The joint neural update executes four coupled steps. First, to guard against overconfident Gaussian log-likelihoods $\widehat\ell_j(a)=2\Rea(a^*\widehat\eta_j)-|a|^2\widehat\epsilon_j$ when the channel estimate is still coarse, a calibration multilayer perceptron (MLP) maps the uncalibrated symbol features $\mathbf{d}_j^{\mathrm{raw}}$ and the mean path curvature $V^{-1}\sum_i \mathbf{k}_i$ to a scalar $c_{\bm{\phi},j}$ that rescales each log-likelihood:
\begin{equation}
 \ell_j^{\mathrm{cal}}(a)=\gamma_j\big[\widehat\ell_j(a)-\max_b\widehat\ell_j(b)\big],\quad
 \gamma_j=e^{3\tanh(c_{\bm{\phi},j}/3)}.
 \label{eq:calibration}
\end{equation}
Replacing the raw log-likelihood entries of $\mathbf{d}_j^{\mathrm{raw}}$ with $\operatorname{asinh}(\ell_j^{\mathrm{cal}}(a))$ yields $\mathbf{d}_{t,j}$. Both $\mathbf{p}_{t,i}$ and $\mathbf{d}_{t,j}$ are layer-normalized, linearly embedded into a common width, and combined with normalized $\mathbf{h}_{p,i},\mathbf{h}_{d,j}$ to form initial embeddings $\mathbf{p}_i^{(0)}$ and $\mathbf{d}_j^{(0)}$, stacked into $\mathbf{P}^{(0)}$ and $\mathbf{D}^{(0)}$.

Second, to extract global multipath patterns from $\mathbf{F}_t$ while preserving continuous localization, a residual convolutional neural network (CNN) with separable delay and Doppler convolutions encodes $\mathbf{F}_t$ into grid features $\mathbf{X}_{\bm{\phi}}$. Bilinear interpolation $\mathcal I_{\bm{\theta}}$ then samples $\mathbf{X}_{\bm{\phi}}$ at the continuous candidate coordinates and produces a continuous gain-prior field $\rho(\bm{\theta})$:
\begin{equation}
 \widetilde{\mathbf{p}}_i=\mathbf{p}_i^{(0)}+\mathbf{W}_f\mathcal I_{\bm{\theta}_i}\mathbf{X}_{\bm{\phi}},\qquad
 \rho(\bm{\theta})=\mathcal I_{\bm{\theta}}(\mathbf{w}_\rho\TT \mathbf{X}_{\bm{\phi}}+b_\rho).
 \label{eq:fieldsampling}
\end{equation}
Here $\rho(\bm{\theta})$ supplies spatial prior information both to the gain-scale update in \eqref{eq:credibility} and to the replacement test in \eqref{eq:replacementtest}.

Third, edge-conditioned graph attention exchanges context between overlapping paths and data symbols \cite{liu2025bigamp}. Specifically, $\widetilde{\mathbf{P}}$ attends over $[\widetilde{\mathbf{P}},\mathbf{D}^{(0)}]$ to form path context $\mathbf{P}^{(1)}$, which in turn updates $\mathbf{D}^{(0)}$ to $\mathbf{D}^{(1)}$. To capture residual inter-symbol interference and global delay--Doppler cues, the symbol branch then applies self-attention \cite{zheng2026} to $\mathbf{D}^{(1)}$, cross-attention from the resulting symbol queries to $\mathbf{X}_{\bm{\phi}}$, and a feed-forward network (FFN) to obtain $\mathbf{D}^+$ with columns $\mathbf{d}_j^+$. Adding the learned logit correction to $\ell_j^{\mathrm{cal}}(a)$ gives the updated symbol probabilities
\begin{equation}
 \pi_j^+(a)=\sm_a\!\left(\ell_j^{\mathrm{cal}}(a)+[\mathbf{W}_c\mathbf{d}_j^++\mathbf{b}_c]_a\right),
 \label{eq:probability}
\end{equation}
from which we compute $\bm{\mu}^+$ and the reconstructed transmit vector $\bar{\mathbf{x}}^+=\mathbf{p}+\mathbf{S}\bm{\mu}^+$ for path refinement.

Finally, so that path regularization and step-size damping reflect the reliability of the newly inferred symbols, each symbol representation is augmented with its posterior-weighted constellation embedding $\mathbf{c}_a$ and fed back to the path branch via symbol-to-path attention $\mathcal G_f$:
\begin{align}
 \overline{\mathbf{d}}_j&=\mathbf{d}_j^++\sum_{a\in\mathcal A}\pi_j^+(a)\mathbf{c}_a,\nonumber\\
 \mathbf{P}^+&=\mathcal G_f(\mathbf{P}^{(1)},\overline{\mathbf{D}};\bm{\mathsf R}),\nonumber\\
 e_i&=\mathbf{w}_e\TT\operatorname{LN}(\mathbf{p}_i^+)+b_e,\nonumber\\
 \bm{\omega}_i&=\spx\!\left(\mathbf{W}_\omega\operatorname{LN}_\omega(\mathbf{p}_i^+)+\mathbf{b}_\omega\right)\in\mathbb R_+^2.
 \label{eq:categoricalfeedback}
\end{align}
Here $\operatorname{LN}$ and $\spx$ denote layer normalization and softplus, the two entries of $\bm{\omega}_i$ govern delay and Doppler damping, respectively, and the updated hidden states are passed to the next layer as $\mathbf{H}_p^+=\mathbf{P}^+$ and $\mathbf{H}_d^+=\mathbf{D}^+$.

\subsection{Regularized Path Refinement}
\label{sec:pathrecovery}
The refinement stage performs regularized Gauss--Newton updates \cite{arjas2026nonlinear}, evaluating each path response to $\bar{\mathbf{x}}^+$ and its derivatives at the current delay and Doppler shift, where $(\mathbf{e},\rho)$ adjust the gain penalties and $\bm{\Omega}$ controls coordinate damping. It first fits the gains at $\mathbf{Q}_t$ and then alternates a coupled coordinate update with a joint gain fit for $J$ steps. Unconstrained coordinates $\mathbf{z}_i=[z_{i,\tau},z_{i,\nu}]\TT$ (stacked into $\mathbf{z}\in\mathbb R^{2V}$) map to the permitted delay--Doppler region as
\begin{equation}
 \tau_i=\tau_{\max}\sig(z_{i,\tau}),\qquad
 \nu_i=\nu_{\max}\tanh(z_{i,\nu}).
 \label{eq:coordinates}
\end{equation}
Here $\sig(u)=(1+e^{-u})^{-1}$ and $\mathbf{C}(\mathbf{z})=\mathbf{C}_{\bm{\mu}^+}(\mathbf{Q}(\mathbf{z}))$. The coordinate vector $\mathbf{z}_t$ at the layer input serves as a fixed reference. With the neural outputs held fixed, the fitting criterion is
\begin{align}
 \mathcal F_t(\mathbf{g},\mathbf{z})={}&\norm{\mathbf{y}-\mathbf{C}(\mathbf{z})\mathbf{g}}^2/\xi_t+\mathbf{g}\HH\bm{\Lambda}_t\mathbf{g}\nonumber\\
 &+\tfrac12(\mathbf{z}-\mathbf{z}_t)\TT \mathbf{R}_z(\mathbf{z}-\mathbf{z}_t).
 \label{eq:objective}
\end{align}
The terms respectively measure prediction error, penalize candidate gains, and penalize coordinate changes from $\mathbf{z}_t$. Here $\bm{\Lambda}_t=\diag(\bm{\lambda})$ contains the gain regularization weights $\lambda_i$. Let $\bm{\zeta}_i=[\partial\tau_i/\partial z_{i,\tau},\partial\nu_i/\partial z_{i,\nu}]\TT$ and let $\bm{\zeta}_{i,t}$ be its value at $\mathbf{z}_t$. The block-diagonal matrix $\mathbf{R}_z$ has blocks
\begin{equation}
 \mathbf{R}_{z,i}=\mathbf{I}_2+\diag\big((\bm{\zeta}_{i,t}\oslash \mathbf{s}_\theta)^2\big),
 \qquad \mathbf{s}_\theta=[0.5,0.25]\TT,
 \label{eq:coordinatepenalty}
\end{equation}
where division $\oslash$ and squaring are elementwise. The scale vector $\mathbf{s}_\theta$ sets the relative units of delay and Doppler displacement. The reconstructed transmit waveform, $\bm{\Lambda}_t$, and $\mathbf{R}_z$ remain fixed during refinement.

Note that \eqref{eq:objective} evaluates the residual at the posterior symbol mean $\bm{\mu}^+$ rather than averaging the squared residual over $\bm{\Pi}^+$. Indeed, for any candidate channel $\mathbf{H}=\mathbf{H}(\mathbf{Q}(\mathbf{z}),\mathbf{g})$ with $\mathbf{B}=\mathbf{H}\mathbf{S}$, the factorized moments in \eqref{eq:moments} yield
\begin{align}
 \E_{\bm{\Pi}^+}\norm{\mathbf{y}-\mathbf{H}(\mathbf{p}+\mathbf{S}\mathbf{d})}^2
 &=\norm{\mathbf{y}-\mathbf{C}(\mathbf{z})\mathbf{g}}^2\nonumber\\
 &\quad+\operatorname{tr}\!\big(\mathbf{B}\diag(\mathbf{v}^+)\mathbf{B}\HH\big).
 \label{eq:expectedfit}
\end{align}

The trace term in \eqref{eq:expectedfit} accounts for symbol uncertainty and vanishes when the symbol variances are zero. Otherwise, it adds a coordinate-dependent, generally non-diagonal quadratic term in the path gains. To avoid recomputing this term during coordinate refinement, \eqref{eq:objective} fits the reconstructed waveform $\bar{\mathbf{x}}^+$ while retaining inter-path coupling in the residual. The symbol probabilities separately inform the gain regularization and coordinate damping. Section~\ref{sec:ablations} reports the NMSE and runtime effects of retaining the trace term in MB-CE.

To stabilize overlapping-path gain estimates, the learned ridge penalty \cite{wu2025trimmed} in our continuous fit combines the previous gain estimate with a neural adjustment for each candidate. Let $s_i=e_i+\rho(\bm{\theta}_i)$ and $E_i=\norm{\mathbf{a}_{\bm{\mu}}(\bm{\theta}_i)}^2/\xi_t$. The gain scale $m_i$ starts at one and is updated at each layer:
\begin{align}
 c_{g,i}&=\big[2(|g_i/m_i|^2+E_i^{-1})\big]^{-1},\nonumber\\
 m_i^+&=\spx\big(\spx^{-1}(m_i)+s_i\big),\nonumber\\
 \lambda_i&=c_{g,i}/(m_i^+)^2.
 \label{eq:credibility}
\end{align}
Here $\spx(u)=\log(1+e^u)$. Equivalently, the gain penalty corresponds to a conditional Gaussian prior $g_i\sim\CN(0,\alpha_i)$ with variance $\alpha_i=1/\lambda_i$. The prior variance depends on the previous-layer gain and gain scale, the inverse response energy, and the learned adjustment. For an isolated response, $E_i^{-1}$ equals the least-squares gain variance under noise variance $\xi_t$, so a positive $s_i$ increases the prior variance and relaxes the gain penalty, whereas a negative $s_i$ strengthens it. Here, the response energy is evaluated using the previous-layer mean $\bm{\mu}$, while the subsequent fit uses the updated mean $\bm{\mu}^+$.

For fixed coordinates, minimizing \eqref{eq:objective} over the gain vector $\mathbf{g}$ gives
\begin{equation}
 \widehat{\mathbf{g}}(\mathbf{z})=(\mathbf{C}(\mathbf{z})\HH \mathbf{C}(\mathbf{z})/\xi_t+\bm{\Lambda}_t)^{-1}\mathbf{C}(\mathbf{z})\HH \mathbf{y}/\xi_t.
 \label{eq:gain}
\end{equation}
For fixed coordinates and penalty weights, this solution is both the mean and mode of the Gaussian posterior on $\mathbf{g}$. For orthogonal response columns, it multiplies each unregularized gain by $E_i^+/(E_i^++\lambda_i)$, where $E_i^+=\norm{\mathbf{a}_{\bm{\mu}^+}(\bm{\theta}_i)}^2/\xi_t$. With overlapping responses, all gains are coupled through $\mathbf{C}\HH \mathbf{C}$. The reconstructed transmit waveform determines the response matrix $\mathbf{C}$ and therefore affects both $\mathbf{C}\HH \mathbf{C}$ and $\mathbf{C}\HH \mathbf{y}$. The penalty matrix $\bm{\Lambda}_t$, constructed separately, remains fixed during the coordinate updates within a layer.

We obtain the response derivatives from \eqref{eq:operator}. Let $s_e^+[m]$ denote the extended IDAFT samples of $\bar{\mathbf{x}}^+$, and set $\mathbf{u}_{\bm{\theta}}=\bm{\mathcal T}(\bm{\theta})\bar{\mathbf{x}}^+$. With $b_\ell=b_\ell(\tau,\nu)$, the derivatives before whitening are
\begin{align}
 [\partial_\tau \mathbf{u}_{\bm{\theta}}]_n
 &=e^{-j2\pi\nu n/N}\sum_{\ell\in\mathcal I}
    (\partial_\tau b_\ell)s_e^+[n-\ell],\nonumber\\
 [\partial_\nu \mathbf{u}_{\bm{\theta}}]_n
 &=e^{-j2\pi\nu n/N}\sum_{\ell\in\mathcal I}
    \left(\partial_\nu b_\ell-\frac{j2\pi n}{N}b_\ell\right)
    s_e^+[n-\ell].
 \label{eq:pulsederivatives}
\end{align}
Applying the fixed whitening matrix $\bm{\mathcal R}$ gives $\partial_\tau \mathbf{a}_{\bm{\mu}^+}(\bm{\theta})$ and $\partial_\nu \mathbf{a}_{\bm{\mu}^+}(\bm{\theta})$. The Doppler derivative includes the dependence of both the pulse coefficient and the phase on Doppler.

Starting from $\mathbf{z}^{(0)}=\mathbf{z}_t$ and $\mathbf{g}^{(0)}=\widehat{\mathbf{g}}(\mathbf{z}_t)$, each Gauss--Newton step linearizes the prediction in the $4V$ real variables $[\Rea g_i,\operatorname{Im}g_i,z_{i,\tau},z_{i,\nu}]_{i=1}^{V}$. At each step, we evaluate the residual $\mathbf{r}=\mathbf{y}-\mathbf{C}(\mathbf{z})\mathbf{g}$, the response $\mathbf{a}_i=\mathbf{a}_{\bm{\mu}^+}(\bm{\theta}_i)$, and the derivatives $\bm{\zeta}_i$ using the current estimates. The Jacobian $\bm{\mathsf J}=[\bm{\mathsf J}_1,\ldots,\bm{\mathsf J}_V]$ gives the following normal matrix and right-hand side:
\begin{align}
 \bm{\mathsf J}_i&=[\mathbf{a}_i,\ j\mathbf{a}_i,\ g_i\partial_\tau \mathbf{a}_i\,\zeta_{i,\tau},\
 g_i\partial_\nu \mathbf{a}_i\,\zeta_{i,\nu}],\nonumber\\
 \bm{\mathcal H}_t&=2\Rea(\bm{\mathsf J}\HH \bm{\mathsf J})/\xi_t+\mathbf{P}_t,\nonumber\\
 \mathbf{b}_t&=2\Rea(\bm{\mathsf J}\HH \mathbf{r})/\xi_t+\mathbf{q}_t.
 \label{eq:direction}
\end{align}
The blocks of $\mathbf{P}_t$ are $2\lambda_i\mathbf{I}_2$ for gains and $\mathbf{R}_{z,i}$ for coordinates. The corresponding blocks of $\mathbf{q}_t$ are $-2\lambda_i[\Rea g_i,\operatorname{Im}g_i]\TT$ and $-\mathbf{R}_{z,i}(\mathbf{z}_i-\mathbf{z}_{t,i})$. The off-diagonal path blocks of the normal matrix couple the path updates. The derivatives with respect to delay and Doppler are scaled by the current complex gains.

We multiply the delay and Doppler entries on the diagonal of $\bm{\mathcal H}_t$ by their respective damping coefficients to form $\mathbf{D}_{\omega,i}$:
\begin{align}
 \mathbf{D}_{\omega,i}&=\diag\!\left(0,0,
 \omega_{i,\tau}[\bm{\mathcal H}_t]_{i\tau,i\tau},
 \omega_{i,\nu}[\bm{\mathcal H}_t]_{i\nu,i\nu}\right),\nonumber\\
 (\bm{\mathcal H}_t+\mathbf{D}_\omega)\bm{\delta}&=\mathbf{b}_t.
 \label{eq:dampeddirection}
\end{align}
Here $\mathbf{D}_\omega$ is block diagonal with blocks $\mathbf{D}_{\omega,i}$, where $[\bm{\mathcal H}_t]_{i\tau,i\tau}$ and $[\bm{\mathcal H}_t]_{i\nu,i\nu}$ denote the diagonal entries of $\bm{\mathcal H}_t$ corresponding to $z_{i,\tau}$ and $z_{i,\nu}$, respectively. The fixed penalty $\mathbf{R}_z$ limits movement from $\mathbf{z}_t$, whereas $\mathbf{D}_\omega$ damps the current increment. While the learned damping $\bm{\Omega}$ remains fixed within the layer, the Jacobian, curvature matrix, and $\mathbf{D}_\omega$ are recomputed at each step. Equivalently, the resulting coordinate step $\bm{\delta}_z$ coincides with eliminating the gain increments via the Schur complement.

The refinement update applies the delay and Doppler components $\bm{\delta}_z^{(j)}$ to update the path coordinates and then re-estimates the gains via \eqref{eq:gain}. Define $\mathbf{D}_z^{(j)}=\diag([\bm{\zeta}_i^{(j)}\oslash \mathbf{s}_\theta]_i)$ using the derivatives at the current iterate and the scale in \eqref{eq:coordinatepenalty}. We rescale the joint coordinate update by a single factor to limit its length without changing its direction:
\begin{align}
 \bm{\Delta}_z^{(j)}&=\frac{\bm{\delta}_z^{(j)}}{\sqrt{1+\norm{\mathbf{D}_z^{(j)}\bm{\delta}_z^{(j)}}_2^2}},\nonumber\\
 \mathbf{z}^{(j+1)}&=\mathbf{z}^{(j)}+\bm{\Delta}_z^{(j)},\qquad
 \mathbf{g}^{(j+1)}=\widehat{\mathbf{g}}(\mathbf{z}^{(j+1)}).
 \label{eq:physicalstep}
\end{align}
After $J$ steps, \eqref{eq:coordinates} yields $\mathbf{Q}^+$ and the final fit yields $\mathbf{g}^+$, from which the channel is reconstructed via \eqref{eq:target} and passed to the next layer together with $\bm{\Pi}^+$ (Section~\ref{sec:replacementschedule}).

\subsection{Inter-Layer Feedback and Path Replacement}
\label{sec:replacementschedule}
After path refinement, we use $(\mathbf{Q}^+,\mathbf{g}^+)$ and $(\bm{\mu}^+,\mathbf{v}^+)$ to compute the residual and per-sample data variance:
\begin{align}
 \mathbf{H}^+&=\mathbf{H}(\mathbf{Q}^+,\mathbf{g}^+),\qquad \mathbf{B}^+=\mathbf{H}^+\mathbf{S},\nonumber\\
 \mathbf{r}^+&=\mathbf{y}-\mathbf{H}^+(\mathbf{p}+\mathbf{S}\bm{\mu}^+),\nonumber\\
 [\mathbf{v}_y^+]_n&=\sum_j|[\mathbf{B}^+]_{n,j}|^2v_j^+.
 \label{eq:predictioncov}
\end{align}
The receiver computes statistics of $\mathbf{r}^+$ and $\mathbf{v}_y^+$ for the noise-variance update. It averages the hidden states over paths and symbols to obtain $\bar{\mathbf{h}}_p,\bar{\mathbf{h}}_d$, then computes
\begin{equation}
 \xi_{t+1}=\xi_t\exp\!\left(2\tanh\mathcal U_{\bm{\phi}}([\bar{\mathbf{h}}_p,\bar{\mathbf{h}}_d,\mathbf{f}_t])\right).
 \label{eq:scalestep}
\end{equation}
Here $\mathbf{f}_t$ contains residual power and predicted data variance, residual spectral statistics, symbol entropy, the mean of $m_i^+$, the current variance $\xi_t$, and layer index $t$. The next layer uses $\xi_{t+1}$ to weight the residual and compute the symbol likelihoods.

For $t<T$, the next layer receives the refined paths $(\mathbf{Q}^+,\mathbf{g}^+)$, symbol probabilities $\bm{\Pi}^+$, gain scales $\mathbf{m}^+$, hidden states $(\mathbf{H}_p^+,\mathbf{H}_d^+)$, and effective variance $\xi_{t+1}$. The next layer also uses the stored correlation field $\rho_t$ when testing path replacements. The GRU in Section~\ref{sec:cavitylearning} is reset to zero at each likelihood evaluation. Following the replacement test below, the receiver rebuilds the channel-dependent matrices, correlations $(u_k,t_k)$, and local curvature features $\mathbf{k}_i$ from the updated estimates.

At the start of each layer $t>1$, the receiver tests whether replacing one current path with a grid point improves a penalized sparse Bayesian marginal-likelihood criterion. We first compute the per-sample data variance $[\mathbf{v}_y]_n=\sum_j|[\mathbf{B}]_{n,j}|^2v_j$ using the current channel and symbol moments. All candidate replacements share the same weight matrix $\mathbf{W}=\diag((\xi_t+\mathbf{v}_y)^{-1/2})$. For either the current or a modified candidate set, let $\widetilde{\mathbf{C}}=\mathbf{W}\mathbf{C}_{\bm{\mu}}(\mathbf{Q})$, $\mathbf{t}_y=\mathbf{W}\mathbf{y}$, $\mathbf{M}=\widetilde{\mathbf{C}}\HH \widetilde{\mathbf{C}}+\bm{\Lambda}_{\mathrm r}$, and $\mathbf{h}=\widetilde{\mathbf{C}}\HH \mathbf{t}_y$. Integrating over $\mathbf{g}\sim\CN(\mathbf{0},\bm{\Lambda}_{\mathrm r}^{-1})$ gives the log marginal likelihood up to a constant:
\begin{equation}
 \mathcal L=-\norm{\mathbf{t}_y}^2+\mathbf{h}\HH \mathbf{M}^{-1}\mathbf{h}-\log\det \mathbf{M}+\log\det\bm{\Lambda}_{\mathrm r}.
 \label{eq:evidence}
\end{equation}
Here $\bm{\Lambda}_{\mathrm r}$ is obtained from \eqref{eq:credibility} using $m_i$ in place of $m_i^+$ and $E_i=\norm{\mathbf{W}\mathbf{a}_{\bm{\mu}}(\bm{\theta}_i)}^2$, and is held fixed across all replacement candidates. Let $\mathcal L_0$ and $\mathcal L_{ik}$ be the log marginal likelihoods for the current set and the set obtained by replacing $\bm{\theta}_i$ with $\bm{\vartheta}_k$. The replacement score is
\begin{align}
 \Delta_{ik}={}&\mathcal L_{ik}-\mathcal L_0
 +\rho_{t-1}(\bm{\vartheta}_k)-\rho_{t-1}(\bm{\theta}_i)-\log(VG).
 \label{eq:replacementtest}
\end{align}

The penalty $-\log(VG)$ accounts for searching over $VG$ possible replacements. If the largest score $\Delta_{ik}$ is positive, the receiver performs that replacement and refits all gains. Otherwise, $(\mathbf{Q},\mathbf{g})$ remains unchanged. Any selected grid point on the search boundary is shifted slightly inward before inverting \eqref{eq:coordinates} to obtain $\mathbf{z}_i$. Upon replacement, the hidden state of the replaced candidate is reset to zero, while the gain scales $\mathbf{m}$ and symbol probabilities $\bm{\Pi}$ remain unchanged. Using a single factorization of the current $\mathbf{M}$ together with Schur complements, the receiver evaluates all replacement scores and computes the updated gains $\mathbf{M}^{-1}\mathbf{h}$ for an accepted replacement, after which it recomputes the channel-dependent inputs and symbol likelihoods in Section~\ref{sec:cavitylearning}.

\subsection{Training and Computational Complexity}
\label{sec:complexity}
Algorithm~\ref{alg:receiver} summarizes LaCE. We train $\bm{\phi}$ end-to-end by minimizing a composite channel-error and symbol cross-entropy loss:
\begin{equation}
 \mathcal J(\bm{\phi})=\E\left[
 \frac{\norm{\widehat{\bm{\Gamma}}-\bm{\Gamma}}_F^2}{r_s\norm{\bm{\Gamma}}_F^2}
 +\lambda_{\mathrm{CE}}\,\mathrm{CE}(\mathbf{d},\bm{\Pi})\right],
 \label{eq:loss}
\end{equation}
where $\mathrm{CE}(\mathbf{d},\bm{\Pi})=-D^{-1}\sum_j\log\pi_j(d_j)$ is the symbol cross-entropy weighted by $\lambda_{\mathrm{CE}}=0.01$, and $r_s$ is a fixed SNR-group scaling factor independent of $\bm{\phi}$ and the predicted channel. During training, gradients backpropagate through both the neural modules and the continuous Gauss--Newton refinement steps, whereas the discrete grid selections in initialization and path replacement are non-differentiable. The same learned noise-variance updates are applied in training and inference.

\begin{algorithm}[!t]
\caption{Proposed LaCE Algorithm}
\label{alg:receiver}
\small
\begin{algorithmic}[1]
\Require Observation $\mathbf{y}$; known $\mathbf{p},\mathbf{S},\mathbf{T}(\bm{\theta}),\mathcal A$, grid and weights $\bm{\phi}$; parameters $V,T,J,K_c$.
\Ensure Channel $\widehat{\bm{\Gamma}}$; symbol probabilities $\bm{\Pi}$.
\State Set $\mathbf{m}\gets\mathbf{1}$, $\pi_j(a)\gets1/|\mathcal A|$, and $\mathbf{H}_p,\mathbf{H}_d\gets\mathbf{0}$.
\State Estimate $\xi_1$ by \eqref{eq:initialscale}.
\State Select $V$ grid points by normalized pilot-residual correlation, refitting all selected gains after each selection.
\State Apply two joint delay--Doppler Gauss--Newton updates with gain re-estimation to initialize $(\mathbf{Q},\mathbf{g},\mathbf{z})$.
\For{$t=1,\ldots,T$}
  \State Compute current symbol moments $(\bm{\mu},\mathbf{v})$ by \eqref{eq:moments} and $\mathbf{B}\gets \mathbf{H}(\mathbf{Q},\mathbf{g})\mathbf{S}$.
  \State \textbf{Path replacement ($t>1$):} If the largest score in \eqref{eq:replacementtest} is positive, replace the corresponding candidate, update its $\mathbf{z}_i$, refit the gains, and reset its hidden state.
  \State Rebuild $\mathbf{B}$ and residual $\mathbf{r}\gets \mathbf{y}-\mathbf{C}_{\bm{\mu}}(\mathbf{Q})\mathbf{g}$.
  \State \textbf{Symbol likelihoods:} Compute likelihoods $(\widehat{\bm{\epsilon}},\widehat{\bm{\eta}})$ by \eqref{eq:pcg}--\eqref{eq:learnedcavity}.
  \State \textbf{Joint neural update:} Form correlation features by \eqref{eq:field} and update $(\mathbf{e},\rho,\bm{\Omega},\bm{\Pi}^+,\mathbf{H}_p^+,\mathbf{H}_d^+)$ from the combined inputs by \eqref{eq:network}.
  \State \textbf{Path refinement:} Form $(\bm{\mu}^+,\mathbf{v}^+)$, $\mathbf{m}^+$ and gain precision $\bm{\Lambda}_t$ by \eqref{eq:moments} and \eqref{eq:credibility}.
  \State Store input coordinates $\mathbf{z}_t\gets \mathbf{z}$ and construct the fixed coordinate penalty $\mathbf{R}_z$, then fit the gains jointly by \eqref{eq:gain}.
  \For{$j=0,\ldots,J-1$}
    \State Solve the coupled damped system \eqref{eq:direction}--\eqref{eq:dampeddirection} at the current $(\mathbf{z},\mathbf{g})$.
    \State Scale the joint coordinate increment and re-estimate the gains by \eqref{eq:physicalstep}.
  \EndFor
  \State Map the final $\mathbf{z}$ to $\mathbf{Q}$ by \eqref{eq:coordinates}, retain the refitted $\mathbf{g}$, and store $\rho_t\gets\rho$.
  \State \textbf{Feedback ($t<T$):} Recompute the residual and data variance by \eqref{eq:predictioncov} and update $\xi_{t+1}$ by \eqref{eq:scalestep}.
  \State $(\bm{\Pi},\mathbf{m},\mathbf{H}_p,\mathbf{H}_d)\gets(\bm{\Pi}^+,\mathbf{m}^+,\mathbf{H}_p^+,\mathbf{H}_d^+)$.
\EndFor
\State Form $\widehat{\bm{\Gamma}}$ from $(\mathbf{Q},\mathbf{g})$ by \eqref{eq:target}, with $P=V$.
\State \Return $\widehat{\bm{\Gamma}},\bm{\Pi}$.
\end{algorithmic}
\end{algorithm}

We now analyze the per-block asymptotic computational complexity and working memory of LaCE, including initialization and all gain and coordinate updates. Pulse-moment tables (evaluated off-grid via cubic Hermite interpolation) and fixed pilot dictionaries are precomputed, while fixed weights, tables, and transforms are excluded from working memory. Let $L$ and $M_p$ denote the numbers of response taps and pulse moments, respectively. Evaluating $U$ responses costs
\begin{equation}
 \mathcal C_a(U)=O\!\left(UL(M_p+N)\right)+\mathcal W_N(U),
 \label{eq:responsecost}
\end{equation}
where $\mathcal W_N(U)=O(N^2U)$ when the receiver applies a dense whitening matrix. Constructing the data channel matrix costs $\mathcal C_B=O(LN^2+N^2\log N)$ using tap summation and chirp--fast Fourier transform (FFT) factorization. The correlation and replacement steps together use $G_u$ distinct grid points. The total cost of waveform synthesis, channel construction, response evaluation, and correlation is
\begin{equation}
 \mathcal C_R=O(ND+LN+LN^2+N^2\log N+NG_u)+\mathcal C_a(U),
 \label{eq:operatorcost}
\end{equation}
per layer, where $U=G_u+O(JV)$ includes the current and updated responses and derivatives. The data channel matrix is constructed at most twice per layer, including the feedback calculation.

Each of the $K_c$ CG steps costs $O(ND)$ for matrix--vector products with $\mathbf{B}$ and $\mathbf{B}\HH$ and $O(D(h_c^2+h_cf_c))$ for the GRU, where $h_c$ and $f_c$ are its hidden and input widths. Including constellation likelihoods gives
\begin{equation}
 \mathcal C_{\mathrm{sym}}=O\!\left(K_cND+K_cD(h_c^2+h_cf_c)+D|\mathcal A|\right).
 \label{eq:symbolcost}
\end{equation}
Joint refinement and gain fits cost $O(J(NV^2+V^3))$. Computing local curvature and received data variance adds $O(NV+ND)$. A shared factorization evaluates all replacement alternatives at cost
\begin{equation}
 \mathcal C_{\mathrm{rep}}=O(NVG+V^2G+NV^2+V^3).
 \label{eq:replacementcost}
\end{equation}

For embedding width $w$ and $E_c$ convolution blocks with filter lengths $k_\tau,k_\nu$, the joint network costs
\begin{align}
 \mathcal C_{\bm{\phi}}=O\!\big(& (E_cG+D+V)w^2+(D+V)|\mathcal A|w\nonumber\\
 &+[E_cG(k_\tau+k_\nu)+V^2+VD]w\nonumber\\
 &+(D^2+DG)w\big).
 \label{eq:neuralcost}
\end{align}
These terms account for feature projections, convolution, graph attention, and self-attention and cross-attention in the symbol branch. The neural working memory is $\mathcal M_{\bm{\phi}}=O((G+D+V)w+(V^2+VD)w+D^2+DG)$, yielding a total per-block working memory of $O(N^2+ND+VLN+V^2+NG_u+\mathcal M_{\bm{\phi}}+Dh_c)$. The initialization cost $\mathcal C_0$ includes pilot-response correlations, sequential grid selection with gain re-estimation, initial channel construction, and two joint coordinate updates. The complete bound is
\begin{align}
 &\mathcal C_{\mathrm{all}}=\mathcal C_0+T\big[\mathcal C_R+\mathcal C_{\mathrm{sym}}+\mathcal C_{\bm{\phi}}\nonumber\\
 &+O(J(NV^2+V^3)+NV+ND)\big]+(T-1)\mathcal C_{\mathrm{rep}}.
 \label{eq:corecost}
\end{align}
In contrast, D-GESBL is dominated by $O(G^3+D^3)$ updates across multiple dictionary-evolution and detection rounds, while pilot-only continuous solvers (MF-GFS, SAGE, and NOMP-CFAR) exhibit data-dependent iteration counts across sweeps or cyclic rounds. The non-learned counterpart MB-CE shares LaCE's fixed outer-round structure but retains $O(D^3)$ inversion for exact symbol likelihoods, whereas LaCE fixes $T$, $K_c$, and $J$ and avoids $O(D^3)$ inversion to achieve tightly bounded per-block complexity.

\section{Simulation Results}
\setlength{\textfloatsep}{6.5pt plus 1.5pt minus 1.5pt}
\setlength{\floatsep}{6pt plus 2pt minus 2pt}
\label{sec:simulations}
\label{sec:simulation}
\subsection{Simulation Setup and Benchmarks}
We evaluate LaCE on 3GPP UMi channels generated by QuaDRiGa \cite{jaeckel2014quadriga} at a carrier frequency of 3.5~GHz and a user velocity of 300~km/h. At each SNR, one AFDM block is simulated per realization over 1,000 independent LOS and 1,000 NLOS channel realizations, where the time-varying coefficients of all QuaDRiGa clusters and subpaths are used to generate the received waveform and reference impulse response while the estimator fits a finite set of continuous delay--Doppler paths.

\begin{table}[!ht]
\caption{Neural Network Architecture}
\label{tab:network_config}
\centering\scriptsize
\setlength{\tabcolsep}{2.5pt}\renewcommand{\arraystretch}{0.96}
\begin{tabular}{@{}p{.28\columnwidth}p{.68\columnwidth}@{}}
\toprule
Component & Architecture and Dimensions \\
\midrule
CG--GRU likelihood & Input/hidden/output $29/48/3$ (11,523 params) \\
Correlation CNN & $12\to96$; 3 separable res.\ blocks ($7\times1$, $1\times5$) \\
Path--symbol attention & Proj.\ $18/40\to96$; 4-head graph attn.\ (QK 24, V 12); shared path FFN $96\to192\to96$; 6-head symbol attn.\ (width 16); symbol FFN $96\to288\to96$ \\
Calibration \& variance & Calib.\ MLP $44\to96\to1$; init $19\to96\to96\to1$; update $200\to1$ \\
Outputs \& total & Gain/field/damping/symbol $1/1/2/16$; total 334,018 \\
\bottomrule
\end{tabular}
\end{table}

The common AFDM frame in \eqref{eq:source}--\eqref{eq:pilotlayout} adopts $N=256$ and $(c_1,c_2)=(3/512,0.003)$, with two pilots placed at bins 110 and 126 using relative phase $\varphi=\pi/2$ and total pilot energy $E_p=0.4N$. The remaining 96 bins in $\mathcal G=\{70,\ldots,167\}$ serve as null guards, while the 158 data bins carry 16-ary quadrature amplitude modulation (16-QAM) symbols with total data energy $DE_s=0.6N$. The sampling rate is 1.92~MHz, and the unit-energy root-raised-cosine (RRC) pulse employs a roll-off factor of 0.25, tapered to zero over $3T_s<|t|<4T_s$. The delay--Doppler search region is $0\le\tau\le12$ and $|\nu|\le1$, which yields $\mathcal I=\{-8,\ldots,20\}$ via \eqref{eq:lagsupport} along with a 20-sample prefix and an 8-sample suffix.

On every realization, the noise-free matched-filter output is normalized to unit average power across the 16-QAM data symbols and the 256 sample positions prior to whitening. Since the unit-energy pulse yields $[\mathbf{C}_q]_{nn}=1$, the pre-whitening SNR is defined as the normalized signal power divided by $\sigma^2$, evaluated from 0 to 30~dB in 5-dB increments under identical channel, data, and noise realizations for all methods. The NMSE of the impulse response in \eqref{eq:target} over $M_b$ test blocks per SNR is defined as
\begin{equation}
 \mathrm{NMSE}_{\mathrm{dB}}=10\log_{10}\!\left[
 \frac1{M_b}\sum_{m=1}^{M_b}
 \frac{\norm{\widehat{\bm{\Gamma}}_m-\bm{\Gamma}_m}_F^2}{\norm{\bm{\Gamma}_m}_F^2}\right],
 \label{eq:nmse}
\end{equation}
evaluated over all taps of the reference impulse response with unestimated taps set to zero. To evaluate the BER, the same linear minimum mean-square error (LMMSE) detector with known noise variance and Gray-labeled 16-QAM is applied to each method's channel estimate.

Since existing learning-based AFDM channel estimators typically assume integer delays \cite{liu2026resnet} or multi-block pilot observations \cite{huang2026afdm_diffusion}, we compare LaCE with four existing single-block estimators and a training-free counterpart developed in this work:
\begin{itemize}
\item \textbf{D-GESBL} \cite{luo2026gesbl} alternates sparse Bayesian channel estimation with data detection. The data decisions are used to rebuild the dictionary for the next round. At each expectation-maximization (EM) step, it refines the delay and Doppler coordinates of up to 47 of the 225 dictionary atoms and updates the noise variance estimate, using up to 40 EM steps per round over eight rounds. We also evaluated its low-complexity generalized approximate message passing variant, D-GAMP-GESBL \cite{luo2026gesbl}, which achieved slightly lower NMSE at low SNR but higher NMSE at high SNR without reducing runtime in paired evaluations, and therefore report D-GESBL in the main comparison.
\item \textbf{MF-GFS} \cite{li2026mf} estimates one path at a time. It uses the largest residual peak for coarse localization, a local grid for fractional delay, and generalized Fibonacci search for fractional Doppler. We retain this procedure and use the joint response of the two pilots under the RRC pulse to score each candidate delay--Doppler location.
\item \textbf{SAGE} \cite{wang2026sage}, based on space-alternating generalized expectation-maximization, updates each path after subtracting the estimated responses of the other paths. We use the single-block variant and supply the true noise variance for its residual and pruning tests.
\item \textbf{NOMP-CFAR} \cite{nomp2016} uses Newtonized orthogonal matching pursuit to add paths at residual correlation peaks, refine their delays and Doppler shifts, and revisit earlier paths, with a constant false-alarm rate (CFAR) test using the true noise variance as the stopping criterion.
\item \textbf{MB-CE} (Model-Based Continuous-parameter Estimator) is a training-free counterpart of LaCE that retains the same pilot initialization, 24 candidates, three alternating rounds, two coupled damped Gauss--Newton steps per round, and SBL path replacement without neural modules. It uses guard-bin quantile noise initialization, exact Gaussian extrinsic soft LMMSE likelihoods, analytical gain penalties in \eqref{eq:credibility} with validation-tuned global scaling and fixed damping, and residual-based noise updates.
\end{itemize}
D-GESBL and MB-CE output both channel and data estimates and initialize the noise variance from the observations, whereas MF-GFS, SAGE, and NOMP-CFAR operate solely on the 50 pilot-region receive bins in $\Omega=\{75,\ldots,124\}$ without data assistance, each subject to a common limit of 24 paths. Specifically, MF-GFS uses eight Fibonacci steps and terminates when the norm of the subtracted response is at most $10^{-3}$ times the pre-subtraction residual norm. SAGE performs up to three passes with twelve sweeps per pass, and NOMP-CFAR executes five cyclic refinement rounds.

Table~\ref{tab:network_config} lists the network dimensions. LaCE uses three estimation layers, 24 path candidates, two initial joint delay--Doppler updates, and two damped Gauss--Newton steps per layer. The joint network in Fig.~\ref{fig:jointnetwork} shares the path FFN between its two path updates and uses a $25\times9$ correlation grid. Each likelihood update uses eight CG steps.

The training set contains 16,800 blocks from 3,360 synthetic channel realizations at 0, 5, 10, 20, and 30~dB, equally divided among sparse, diffuse, and clustered channels. For the QuaDRiGa evaluations, LaCE is trained from random initialization for 4,760 AdamW iterations with a batch size of 10, and checkpoints are selected on a disjoint validation set of 96 synthetic channels according to the worst-case SNR-weighted NMSE across channel families and SNR groups, using the mean validation loss to break ties.

\subsection{Performance on UMi LOS Channels}
\begin{figure}[!tbp]
\centering\includegraphics[width=0.84\columnwidth]{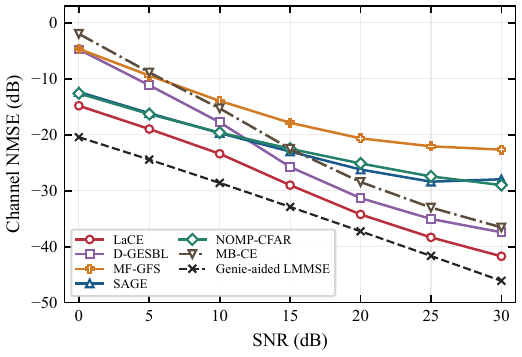}
\caption{Channel estimation NMSE on QuaDRiGa UMi LOS channels.}
\label{fig:nmse}
\end{figure}

Figure~\ref{fig:nmse} compares the channel estimation NMSE on the UMi LOS realizations. From 0 to 10~dB, LaCE outperforms the best benchmark, namely NOMP-CFAR at 0 and 5~dB and SAGE at 10~dB, by 2.21 to 3.70~dB. At higher SNR, D-GESBL benefits from more reliable data decisions but remains 4.29~dB above LaCE at 30~dB. The gap between MB-CE and LaCE ranges from 5.16 to 12.86~dB and is widest at low SNR, where symbol errors affect data-aided refinement. In contrast, NOMP-CFAR improves little at high SNR, SAGE degrades as data leakage becomes dominant, and MF-GFS has the highest NMSE at 30~dB because it does not revisit earlier path estimates. LaCE attains the lowest NMSE at every SNR, within 3.01 to 5.61~dB of the genie-aided LMMSE bound\textsuperscript{\cite{luo2026gesbl}}. This bound averages $\operatorname{tr}[\mathbf{G}\HH\mathbf{G}(\mathbf{A}\HH\mathbf{A}/\sigma^2+\mathbf{I})^{-1}]/\norm{\bm{\Gamma}}_F^2$, where $\mathbf{G}$ contains the vectorized true cluster responses including their amplitudes, $\mathbf{A}$ contains their whitened receive responses to the known transmitted block, and the cluster coefficients have independent $\CN(0,1)$ priors.

\begin{figure}[!t]
\centering\includegraphics[width=0.84\columnwidth]{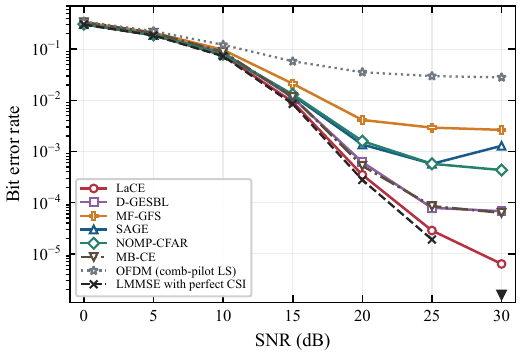}
\caption{UMi LOS BER with common LMMSE detection (downward triangle: zero errors with perfect CSI).}
\label{fig:ber}
\end{figure}

\begin{figure}[!t]
\centering\includegraphics[width=0.84\columnwidth]{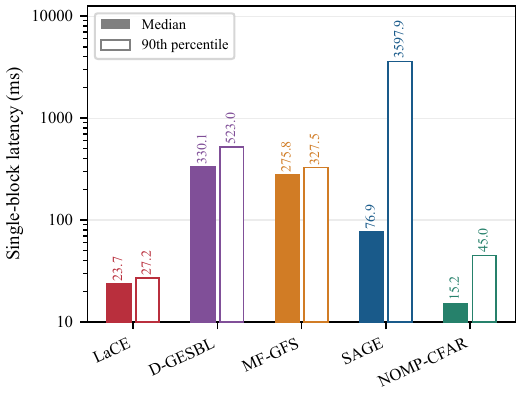}
\caption{Median and 90th-percentile estimation latency over 100 UMi LOS blocks. MB-CE was timed separately.}
\label{fig:latency}
\end{figure}

Figure~\ref{fig:ber} reports the corresponding BER performance together with the perfect-CSI reference obtained under the same LMMSE detector. From 0 to 10~dB, LaCE closely tracks this reference, and NOMP-CFAR and SAGE achieve comparable BERs despite the clear NMSE gaps observed in Fig.~\ref{fig:nmse}. As the SNR increases, the BER curves diverge noticeably. Although D-GESBL and MB-CE are the closest benchmarks from 15 to 30~dB, LaCE remains substantially closer to the perfect-CSI reference and achieves the lowest BER among all six AFDM estimators at every considered SNR. By contrast, the comb-pilot OFDM baseline with 64 pilots, least-squares (LS) estimation, and single-tap equalization exhibits an error floor near $2.8\times 10^{-2}$ due to severe Doppler-induced intercarrier interference.

Figure~\ref{fig:latency} compares the per-block estimation latency of LaCE and four existing benchmarks over 100 LOS blocks on an RTX~5060 GPU. Timing includes block initialization and data transfers and excludes offline setup and the final LMMSE detection. NOMP-CFAR has the lowest median latency, whereas LaCE has the lowest 90th-percentile latency among the methods in Fig.~\ref{fig:latency} (27.2~ms). SAGE is faster than MF-GFS at the median but has a much higher 90th-percentile latency. Although D-GESBL achieves the closest high-SNR NMSE among the benchmarks, it requires more than ten times the runtime of LaCE at both percentiles. In a separate paired timing evaluation under the same configuration, MB-CE achieves a 90th-percentile latency of 20.40~ms compared with 22.49~ms for LaCE, running about 2~ms faster per block due to the absence of neural updates.

\subsection{Convergence, Ablations, and Generalization}
\label{sec:ablations}
\begin{figure}[!tbp]
\centering\includegraphics[width=0.80\columnwidth]{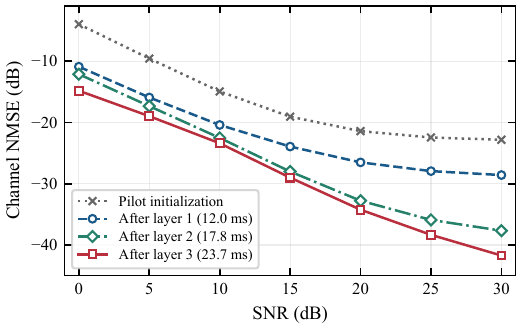}
\caption{NMSE after pilot initialization and each layer over 1,000 LOS realizations, with cumulative median runtimes over 100 blocks.}
\label{fig:layerdepth}
\end{figure}

Figure~\ref{fig:layerdepth} compares the pilot-based initialization with the estimate after each refinement layer. At low SNR, the first layer provides the largest improvement, whereas the first two layers account for most of the gain at intermediate SNRs. The initialization and first-layer curves flatten at the highest SNRs, but the later layers continue to lower the NMSE. At 30~dB, the third layer alone improves on the second by about 4~dB. The loss from stopping after an early layer is therefore most pronounced at high SNR, where successive rounds of symbol inference and path refinement yield further gains.

\begin{figure}[!tbp]
\centering\includegraphics[width=0.84\columnwidth]{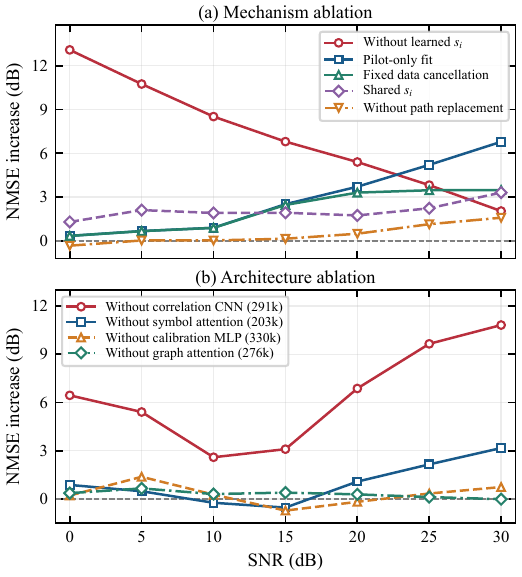}
\caption{Ablation study on 100 UMi LOS realizations: (a) mechanism ablation using the trained LaCE parameters; (b) architecture ablation under 1,120 training iterations.}
\label{fig:ablations}
\end{figure}

Figure~\ref{fig:ablations}(a) separates the effects of joint fitting, gain regularization, and path replacement. Pilot-only fitting and fixed data cancellation perform similarly to joint fitting at low SNR, but both suffer noticeable degradation as the SNR increases. Although fixed cancellation recovers part of the high-SNR gain, the remaining gap confirms the benefit of jointly fitting the pilot and reconstructed data responses. Setting the learned adjustment $s_i$ to zero increases the NMSE, especially at low SNR. Using a common adjustment across candidates recovers much of the loss at low SNR, while path-specific adjustments yield lower NMSE at high SNR. Path replacement also improves the high-SNR estimates at the cost of 2.6~ms per block.

The architecture ablations in Fig.~\ref{fig:ablations}(b) use the same training set over 1,120 training iterations. Removing the correlation CNN yields the largest NMSE degradation across all SNRs. Symbol attention contributes most at high SNR, whereas removing either graph attention or the calibration MLP has a smaller impact on the NMSE. In addition, CG--GRU reduces the median GPU likelihood runtime by about 53\% relative to exact $\bm{\mathsf K}^{-1}$ diagonals with under 0.09~dB NMSE variation. Retaining the trace term of \eqref{eq:expectedfit} in MB-CE improves its NMSE at low SNR with little gain at higher SNR, but increases the CPU runtime by a factor of 9.2 while remaining inferior to LaCE.

\begin{figure}[!tbp]
\centering\includegraphics[width=0.84\columnwidth]{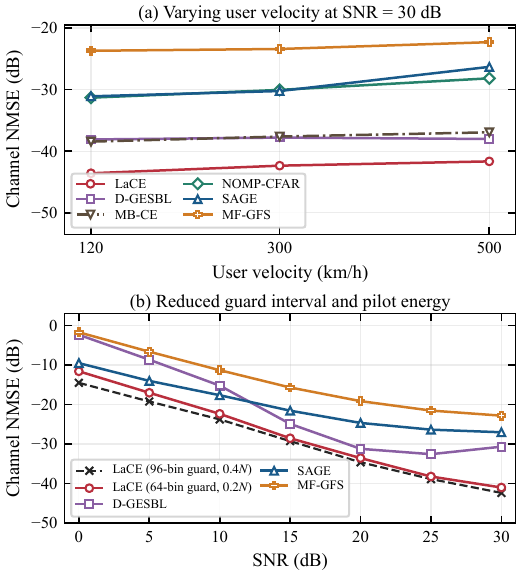}
\caption{Robustness on 100 UMi LOS realizations without retraining: (a) channel NMSE versus user velocity at $\mathrm{SNR}=30$~dB; (b) channel NMSE with a 64-bin guard interval and $E_p=0.2N$.}
\label{fig:robustness}
\end{figure}

\begin{table}[!t]
\caption{Channel NMSE and BER on UMi NLOS Channels}
\label{tab:nlos}
\centering\scriptsize
\setlength{\tabcolsep}{2.1pt}
\begin{tabular}{@{}rrrrr@{\hspace{5pt}}rrrr@{}}
\toprule
 & \multicolumn{4}{c}{NMSE (dB)} & \multicolumn{4}{c}{BER ($\times 10^{-3}$)} \\
\cmidrule(lr){2-5}\cmidrule(l){6-9}
SNR & LaCE & D-GESBL & NOMP & MB-CE & LaCE & D-GESBL & NOMP & MB-CE \\
\midrule
\NlosExpandedRows
\bottomrule
\end{tabular}
\end{table}

Figure~\ref{fig:robustness} evaluates the robustness of LaCE to varying user velocity and reduced pilot and guard overhead without retraining. In Fig.~\ref{fig:robustness}(a), the NMSE of LaCE degrades only slightly as the user velocity increases from 120 to 500~km/h, whereas SAGE and NOMP-CFAR exhibit more pronounced degradation. LaCE maintains an NMSE gain of at least 3.6~dB over D-GESBL across all considered velocities. In Fig.~\ref{fig:robustness}(b), reducing the guard interval to 64 null bins allocates 74.2\% of the subcarriers to data, while the pilot energy is halved to $E_p=0.2N$. Although the reduced overhead has a larger impact at low SNR, LaCE outperforms all benchmarks across the considered SNR range and retains a clear high-SNR advantage over D-GESBL.

Table~\ref{tab:nlos} reports the results on UMi NLOS channels with the same AFDM frame and LMMSE detector, without retraining or parameter retuning. The relative performance of the benchmarks varies with the SNR, with NOMP-CFAR performing best at low SNR and D-GESBL and MB-CE achieving higher accuracy at high SNR. By contrast, LaCE attains the lowest NMSE and BER across the entire SNR range. At 30~dB, it retains a 3.54~dB NMSE advantage over D-GESBL, the closest benchmark in NMSE, and also yields a lower BER.

\section{Conclusion}
\label{sec:conclusion}
In this paper, we proposed LaCE, a learning-assisted continuous channel estimator for AFDM systems with fractional delay and Doppler. Starting with a pilot-based initialization, LaCE alternates between CG--GRU symbol inference and coupled continuous multipath refinement, jointly fitting the embedded pilot and reconstructed data responses via damped Gauss--Newton updates and regularized least squares. Learned likelihood calibration, path-specific gain regularization, and damping further suppress data-feedback error propagation. Simulation results on 3GPP UMi LOS and NLOS channels demonstrated consistent NMSE and BER gains over representative sparse-recovery, Bayesian, and continuous-parameter benchmarks across the entire SNR range, with low estimation latency, while ablation studies verified the roles of joint fitting, path-specific regularization, and recurrent likelihood approximation.

\bibliographystyle{IEEEtran}
\bibliography{references}
\end{document}